\documentclass[12pt,a4paper,notoc]{article}
\pdfoutput=1
 
\usepackage{jheppub-arxiv}
\usepackage[usenames,dvipsnames,table]{xcolor}
 
\usepackage{amssymb} 
\usepackage{amsmath}
\usepackage{mathtools}
\usepackage{amsfonts}    
\usepackage{dsfont}
\usepackage{pdfpages}
\usepackage{verbatim}
\usepackage{stmaryrd}
\usepackage{graphicx}
\usepackage{import}
\usepackage{etoc}
\usepackage{enumitem}
\usepackage{tensor}
\usepackage{mathrsfs}
\usepackage{textgreek} 
\usepackage[mathscr]{euscript}
\usepackage[smalltableaux]{ytableau}
\ytableausetup{centertableaux}
\usepackage{tikz}
\usetikzlibrary{decorations.pathreplacing, decorations.markings,calc,shapes.misc,decorations.pathmorphing,patterns.meta, math,positioning}
\usepackage{tabularx,ragged2e}
\usepackage{physics}

\usepackage{slashed}
\usepackage{datetime}
\usepackage{hyperref}
\hypersetup{
    pdfencoding=unicode,
	colorlinks=true,
	urlcolor=Maroon,
	linkcolor=RoyalBlue,
	citecolor=Maroon,
	pdftitle={Records from the S-Matrix Marathon: Tasty Bits of Several Complex Variables},
	pdfauthor={},
	pdfdisplaydoctitle=true,
	pdfstartview=FitH,
	linktocpage=true
}

\usetikzlibrary{shapes}
\usepackage{enumitem}
\usepackage[export]{adjustbox}
\usepackage{nccmath}
\usepackage{bbm}
\usepackage{braket}
\usepackage{empheq}
\usepackage{cancel}
\usepackage{cleveref}
\usepackage[most]{tcolorbox}

\definecolor{charcoal}{HTML}{343837}

\usepackage{bm}
\definecolor{yellowish}{rgb}{0.880722,0.611041,0.142051}

\newcommand{\ba}{\begin{align}}

\newcommand{\be}{\begin{equation}}
\newcommand{\ee}{\end{equation}}
\def\bd{\begin{tikzpicture}}
\def\ed{\end{tikzpicture}}
\renewcommand{\abs}[1]{\left| #1 \right|}

\renewcommand\Im{\mathop{\text{Im}}}

\renewcommand\Re{\mathop{\text{Re}}}

\allowdisplaybreaks

\def\XXint#1#2#3{{\setbox0=\hbox{$#1{#2#3}{\int}$}
     \vcenter{\hbox{$#2#3$}}\kern-.5\wd0}}

\definecolor{light-gray}{gray}{0.75}

\newcommand\CP{\mathbb{CP}}

\renewcommand\d{\text{d}}

\newcommand{\e}{\mathrm{e}}

\newcommand{\R}{\mathrm{R}}

\newcommand{\D}{\mathbb{D}}

\newcommand{\C}{\mathbb{C}}
\renewcommand{\R}{\mathbb{R}}

\newcommand{\B}{\mathbb{B}}
\newcommand{\N}{\mathbb{N}}

\renewcommand{\leq}{\leqslant}
\renewcommand{\geq}{\geqslant}

\definecolor{dpurple}  {RGB} {189,  147,  249}

\usepackage{ntheorem}
\usepackage[framemethod=tikz,footnoteinside=false,hidealllines=true,topline=true,bottomline=true,linecolor=black!10!white,linewidth=1pt,innerleftmargin=0em,innerrightmargin=0em,frametitlerule=true,ntheorem]{mdframed}
\theorembodyfont{\upshape}

\newmdtheoremenv{mtheorem}{Theorem}[section]
\newmdtheoremenv[]{mdexample}{Example}[section]
\newmdtheoremenv{mdremark}{Remark}[section]
\newmdtheoremenv{mddefinition}{Definition}[section]
\newmdtheoremenv{mdcorollary}{Corollary}[section]
\newmdtheoremenv{mdproposition}{Proposition}[section]
\newmdtheoremenv{QA}{Audience question}[section]

\title{{\Large\normalfont Records from the S-Matrix Marathon:}\\ Tasty Bits of Several Complex Variables}

\author{{\normalfont Lecturers:}}
\author[1]{Sean~Curry,}
\author[1]{Jiří~Lebl}

\author{\\ {\normalfont Notes written by:}}
\author[2]{Mathieu~Giroux,}
\author[3]{Holmfridur~S.~Hannesdottir,}
\author[3,4,5]{Sebastian~Mizera,}
\author[2]{Celina~Pasiecznik}

\affiliation[1]{Departemento pri Matematiko de Oklahoma {\^S}tata Universitato,\\ Stillwater, OK 74078, Usono}
\affiliation[2]{Department of Physics, McGill University, 3600 Rue University,\\ Montr\'eal, H3A 2T8, QC Canada}
\affiliation[3]{Institute for Advanced Study, Princeton, NJ 08540, USA}
\affiliation[4]{Department of Physics, Princeton University, Princeton, NJ 08544, USA}
\affiliation[5]{Princeton Center for Theoretical Science,\\ Princeton University, Princeton, NJ 08544, USA}

\abstract{We will cover the basics of several complex variables in 4 lectures: Basic properties of holomorphic functions in several variables, the notion of pseudoconvexity, CR functions and CR geometry, and the $\bar\partial$-problem.  The main underlying idea is to connect various characterizations of domains of holomorphy, that is, the natural domains of definition for holomorphic functions.  In the process we will connect the function theory on the domain to geometric properties of the boundary, and discuss the relationship between the boundary values and the functions themselves and extension of holomorphic functions from subspaces.

These notes are based on a series of lectures held during the S-Matrix Marathon workshop at the Institute for Advanced Study on 11--22 March 2024.
}

\begin{document}

\theoremstyle{definition}
\newtheorem{definition}{Definition}[section]
\theoremstyle{remark}
\newtheorem*{remark}{Remark}

\setcounter{tocdepth}{2}
\maketitle
\setcounter{page}{1}

\setcounter{tocdepth}{4}
\setcounter{secnumdepth}{4}

\makeatletter
\g@addto@macro\bfseries{\boldmath}
\makeatother

\newpage
\section*{Preface}

This article is a chapter from the \emph{Records from the S-Matrix Marathon}, a series of lecture notes covering selected topics on scattering amplitudes~\cite{RecordsBook}. They are based on lectures delivered during a workshop on 11--22 March 2024 at the Institute for Advanced Study in Princeton, NJ. We hope that they can serve as a pedagogical introduction to the topics surrounding the S-matrix theory.

These lecture notes were prepared by the above-mentioned note-writers in collaboration with the lecturers.

\vfill
\section*{Acknowledgments}

S.C. and J.L. would like to express their thanks to the organizers for the wonderful conference (and for bringing in mathematicians)! S.C. is supported in part by the Simons Foundation, grant MPS-TSM-00002876. J.L. is supported in part by the Simons Foundation collaboration grant 710294. M.G.’s and C.P.'s work is supported in parts by the National Science and Engineering Council of Canada (NSERC) and the Canada Research
Chair program, reference number CRC-2022-00421. Additionally, C.P. is supported by the Walter C. Sumner Memorial Fellowship.
H.S.H. gratefully acknowledges funding provided by the J. Robert Oppenheimer Endowed Fund of the Institute for Advanced Study. 
S.M. gratefully acknowledges funding provided by the Sivian Fund and the Roger Dashen Member Fund at the Institute for Advanced Study. 
This material is based upon work supported by the U.S. Department of Energy, Office of Science, Office of High Energy Physics under Award Number DE-SC0009988.

The S-Matrix Marathon workshop was sponsored by the Institute for Advanced Study and the Carl P. Feinberg Program in Cross-Disciplinary Innovation.

\setcounter{section}{1}
\newcommand{\esssup}{\operatorname{ess~sup}}
\newcommand{\essran}{\operatorname{essran}}
\newcommand{\innprod}[2]{\langle #1 | #2 \rangle}
\newcommand{\linnprod}[2]{\langle #1 , #2 \rangle}
\newcommand{\supp}{\operatorname{supp}}
\newcommand{\Nul}{\operatorname{Nul}}
\newcommand{\Ran}{\operatorname{Ran}}
\renewcommand{\abs}[1]{\left\lvert {#1} \right\rvert}
\renewcommand{\norm}[1]{\left\lVert {#1} \right\rVert}
\newcommand{\sabs}[1]{\lvert {#1} \rvert}
\newcommand{\snorm}[1]{\lVert {#1} \rVert}
\newcommand{\babs}[1]{\bigl\lvert {#1} \bigr\rvert}
\newcommand{\bnorm}[1]{\bigl\lVert {#1} \bigr\rVert}
\newcommand{\Babs}[1]{\Bigl\lvert {#1} \Bigr\rvert}
\newcommand{\Bnorm}[1]{\Bigl\lVert {#1} \Bigr\rVert}
\newcommand{\bbabs}[1]{\biggl\lvert {#1} \biggr\rvert}
\newcommand{\bbnorm}[1]{\biggl\lVert {#1} \biggr\rVert}
\newcommand{\BBabs}[1]{\Biggl\lvert {#1} \Biggr\rvert}
\newcommand{\BBnorm}[1]{\Biggl\lVert {#1} \Biggr\rVert}

\renewcommand{\C}{{\mathbb{C}}}
\renewcommand{\R}{{\mathbb{R}}}
\newcommand{\Z}{{\mathbb{Z}}}
\renewcommand{\N}{{\mathbb{N}}}
\newcommand{\Q}{{\mathbb{Q}}}
\renewcommand{\D}{{\mathbb{D}}}
\newcommand{\F}{{\mathbb{F}}}

\newcommand{\bB}{{\mathbb{B}}}
\newcommand{\bC}{{\mathbb{C}}}
\newcommand{\bR}{{\mathbb{R}}}
\newcommand{\bZ}{{\mathbb{Z}}}
\newcommand{\bN}{{\mathbb{N}}}
\newcommand{\bQ}{{\mathbb{Q}}}
\newcommand{\bD}{{\mathbb{D}}}
\newcommand{\bF}{{\mathbb{F}}}
\newcommand{\bH}{{\mathbb{H}}}
\newcommand{\bO}{{\mathbb{O}}}
\newcommand{\bP}{{\mathbb{P}}}
\newcommand{\bK}{{\mathbb{K}}}
\newcommand{\bV}{{\mathbb{V}}}
\renewcommand{\CP}{{\mathbb{CP}}}
\newcommand{\RP}{{\mathbb{RP}}}
\newcommand{\HP}{{\mathbb{HP}}}
\newcommand{\OP}{{\mathbb{OP}}}
\newcommand{\sA}{{\mathscr{A}}}
\newcommand{\sB}{{\mathscr{B}}}
\newcommand{\sC}{{\mathscr{C}}}
\newcommand{\sF}{{\mathscr{F}}}
\newcommand{\sG}{{\mathscr{G}}}
\newcommand{\sH}{{\mathscr{H}}}
\newcommand{\sM}{{\mathscr{M}}}
\newcommand{\sO}{{\mathscr{O}}}
\newcommand{\sP}{{\mathscr{P}}}
\newcommand{\sQ}{{\mathscr{Q}}}
\newcommand{\sR}{{\mathscr{R}}}
\newcommand{\sS}{{\mathscr{S}}}
\newcommand{\sI}{{\mathscr{I}}}
\newcommand{\sL}{{\mathscr{L}}}
\newcommand{\sK}{{\mathscr{K}}}
\newcommand{\sU}{{\mathscr{U}}}
\newcommand{\sV}{{\mathscr{V}}}
\newcommand{\sX}{{\mathscr{X}}}
\newcommand{\sY}{{\mathscr{Y}}}
\newcommand{\sZ}{{\mathscr{Z}}}
\newcommand{\fS}{{\mathfrak{S}}}

%
%
\makeatletter
\let\save@mathaccent\mathaccent
\newcommand*\if@single[3]{%
  \setbox0\hbox{${\mathaccent"0362{#1}}^H$}%
  \setbox2\hbox{${\mathaccent"0362{\kern0pt#1}}^H$}%
  \ifdim\ht0=\ht2 #3\else #2\fi
  }
\newcommand*\rel@kern[1]{\kern#1\dimexpr\macc@kerna}
\newcommand*\widebar[1]{\@ifnextchar^{{\wide@bar{#1}{0}}}{\wide@bar{#1}{1}}}
\newcommand*\wide@bar[2]{\if@single{#1}{\wide@bar@{#1}{#2}{1}}{\wide@bar@{#1}{#2}{2}}}
\newcommand*\wide@bar@[3]{%
  \begingroup
  \def\mathaccent##1##2{%
    \let\mathaccent\save@mathaccent
    \if#32 \let\macc@nucleus\first@char \fi
    \setbox\z@\hbox{$\macc@style{\macc@nucleus}_{}$}%
    \setbox\tw@\hbox{$\macc@style{\macc@nucleus}{}_{}$}%
    \dimen@\wd\tw@
    \advance\dimen@-\wd\z@
    \divide\dimen@ 3
    \@tempdima\wd\tw@
    \advance\@tempdima-\scriptspace
    \divide\@tempdima 10
    \advance\dimen@-\@tempdima
    \ifdim\dimen@>\z@ \dimen@0pt\fi
    \rel@kern{0.6}\kern-\dimen@
    \if#31
      \overline{\rel@kern{-0.6}\kern\dimen@\macc@nucleus\rel@kern{0.4}\kern\dimen@}%
      \advance\dimen@0.4\dimexpr\macc@kerna
      \let\final@kern#2%
      \ifdim\dimen@<\z@ \let\final@kern1\fi
      \if\final@kern1 \kern-\dimen@\fi
    \else
      \overline{\rel@kern{-0.6}\kern\dimen@#1}%
    \fi
  }%
  \macc@depth\@ne
  \let\math@bgroup\@empty \let\math@egroup\macc@set@skewchar
  \mathsurround\z@ \frozen@everymath{\mathgroup\macc@group\relax}%
  \macc@set@skewchar\relax
  \let\mathaccentV\macc@nested@a
  \if#31
    \macc@nested@a\relax111{#1}%
  \else
    \def\gobble@till@marker##1\endmarker{}%
    \futurelet\first@char\gobble@till@marker#1\endmarker
    \ifcat\noexpand\first@char A\else
      \def\first@char{}%
    \fi
    \macc@nested@a\relax111{\first@char}%
  \fi
  \endgroup
}
\makeatother

\newpage

\section*{\label{ch:LeblCurry}Tasty Bits of Several Complex Variables\\
\normalfont{\textit{Sean Curry, Jiří Lebl}}}

\setcounter{section}{0}

\noindent\rule{\textwidth}{0.25pt}
\vspace{-0.8em}
\etocsettocstyle{\noindent\textbf{Contents}\vskip0pt}{}
\localtableofcontents
\vspace{0.5em}
\noindent\rule{\textwidth}{0.25pt}
\vspace{1em}

\noindent
These lecture notes roughly correspond to the first 4 chapters of the book of the same name \cite{lebl2019tasty}.

\section[Holomorphic functions in several variables]{Holomorphic functions in several variables\\
\normalfont{\textit{Jiří Lebl}}}

\subsection{Onto several variables}

Let $\C^n$ be the complex Euclidean space with coordinates
$z = (z_1,z_2,\ldots,z_n) \in \C^n$ and it will be useful to treat it as two copies of the real space,
$\C^n \cong \R^n \times \R^n = \R^{2n}$
with
\be
z = x+iy, \qquad \bar{z}=x-iy, \qquad x,y \in \R^n, \qquad i = \sqrt{-1} .
\ee
We call $z$ the \emph{holomorphic coordinates} and $\bar{z}$ \emph{antiholomorphic coordinates}.
Let us define a \emph{polydisc} $\Delta_\rho(a)$
with \emph{polyradius} $\rho = (\rho_1,\rho_2,\ldots,\rho_n)$
and
\emph{center} $a \in \C^n$ as
\be
\Delta_\rho(a)  \overset{\text{def}}{=}
\bigl\{ z \in \C^n : \sabs{z_k - a_k} < \rho_k ~\text{for $k=1,2,\ldots,n$} \bigr\} .
\ee
(If $\rho$ is a number, we mean $\rho_k = \rho$ for all $k$.)
In two variables, a polydisc is sometimes called a \emph{bidisc}. In particular, the \emph{unit polydisc} is given by
is
\be
\D^n = \D \times \D \times \cdots \times \D = \Delta_1(0).
\ee 

We will use the following notation for the Euclidean inner product on $\C^n$ and the standard Euclidean norm on $\C^n$:
\be
\langle z,w \rangle = z \cdot \bar{w}\, ,
\qquad
\snorm{z} = \sqrt{\langle z , z \rangle}.
\ee
Then we define $B_\rho(a)$ to be the ball in the metric $\norm{\cdot}$. For example, $\bB_n = B_1(0)$ is the \emph{unit ball}.

\begin{mdexample}
In more than one complex dimension, it is difficult to draw pictures because we lack dimensions on our paper. We instead draw pictures by plotting against the modulus of the variables.
For example, the unit polydisc $\D^2$ and unit ball $\B_2$ in $n=2$ complex dimensions can be visualized as follows:
\be
\begin{adjustbox}{valign=c}
\scalebox{1}{
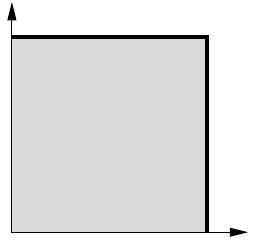
}
\qquad\qquad
\scalebox{1}{
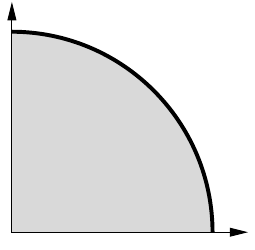
}
\end{adjustbox}
\ee
\end{mdexample}
Not every domain (by domain we mean an open connected set) can be drawn like this. If it can, it is called a \emph{Reinherdt domain}.

The function $f \colon U \subset \C^n \to \C$ is called \emph{holomorphic} if $f$
is complex differentiable in each variable separately, that is, if
\begin{equation}
z_\ell \mapsto f(z_1,\ldots,z_\ell,\ldots,z_n)
\text{ is complex differentiable for every } \ell\,.
\end{equation}
Let us write $\sO(U)$ to denote the set of holomorphic functions on $U$. Here, the letter $\sO$ is used to recognize the fundamental contribution to several complex variables by Kiyoshi Oka.

From now on we will benefit from using the language of differential forms. Exterior derivative leads to $1$-forms
\be
\d z_\ell = \d x_\ell + i\, \d y_\ell, \qquad
\d\bar{z}_\ell = \d x_\ell - i\, \d y_\ell .
\ee
As in one variable, we define the \emph{Wirtinger operators}:
\begin{equation}
\frac{\partial}{\partial z_\ell}  \overset{\text{def}}{=}
\frac{1}{2} \left(
\frac{\partial}{\partial x_\ell} - i \frac{\partial}{\partial y_\ell}
\right) ,
\qquad
\frac{\partial}{\partial \bar{z}_\ell}  \overset{\text{def}}{=}
\frac{1}{2} \left(
\frac{\partial}{\partial x_\ell} + i \frac{\partial}{\partial y_\ell}
\right) .
\end{equation}
These are determined by being the dual bases of $\d z$ and $\d\bar{z}$
\be
\d z_k\left(\frac{\partial}{\partial z_\ell}\right) = \delta_\ell^k, \quad
\d z_k\left(\frac{\partial}{\partial \bar{z}_\ell}\right) = 0, \quad
\d\bar{z}_k\left(\frac{\partial}{\partial z_\ell}\right) = 0, \quad
\d\bar{z}_k\left(\frac{\partial}{\partial \bar{z}_\ell}\right) = \delta_\ell^k\, .
\ee
Alternatively, we might have said that $f$ is holomorphic if it satisfies the \emph{Cauchy--Riemann (CR) equations}:
\begin{equation}
\frac{\partial f}{\partial \bar{z}_\ell}  = 0 \quad \text{for $\ell=1,2,\ldots,n$}\, .
\end{equation}
If $f$ is holomorphic, then its derivatives can be obtained by taking limits
\begin{equation}
\frac{\partial f}{\partial z_k}(z)
=
\lim_{\xi \in \C \to 0} \frac{f(z_1,\ldots,z_k+\xi,\ldots,z_n) - f(z)}{\xi}\,
.
\end{equation}
From now on, we will write a smooth function $f \colon U \subset \C^n \to \C$ simply
as $f(z,\bar{z})$.

\begin{mdexample}
If $f$ is a polynomial (in $x$ and $y$), write
\be
x = \dfrac{z+\bar{z}}{2}, \qquad
y = \dfrac{z-\bar{z}}{2i}\,,
\ee
and it really does become a polynomial in $z$ and $\bar{z}$.
E.g.,
\be
2x_1 +2y_1 + 4y_2^2 = 
(1-i)z_1+(1+i)\bar{z}_1-z_2^2+2z_2\bar{z}_2 - \bar{z}_2^2 .
\ee
$f$ is holomorphic if
it does not depend on $\bar{z}$.
\end{mdexample}

The derivatives satisfy the chain rule. Suppose
$f \colon U \subset \C^n \to V \subset \C^m$, $g \colon V \to \C$
and that the variables are $z \in \C^n$ and $w \in \C^m$. Then
\begin{align}
\frac{\partial}{\partial z_\ell} \left[ g \circ f \right]
& =
\sum_{k=1}^m \bigg(
\frac{\partial g}{\partial w_k}
\frac{\partial f_k}{\partial z_\ell}
+
\cancel{\frac{\partial g}{\partial \bar{w}_k}
\frac{\partial \bar{f}_k}{\partial z_\ell}
}\bigg) ,
\\
\frac{\partial}{\partial \bar{z}_\ell} \left[ g \circ f \right]
& =
\sum_{k=1}^m 
\cancel{\left(
\frac{\partial g}{\partial w_k}
\frac{\partial f_k}{\partial \bar{z}_\ell}
+
\frac{\partial g}{\partial \bar{w}_k}
\frac{\partial \bar{f}_k}{\partial \bar{z}_\ell}
\right)} = 0\, ,
\end{align}
provided that $g$ and $f$ are holomorphic.

\begin{mtheorem}[Cauchy integral formula]
Let $\Delta \subset \C^n$ be a polydisc.
Suppose
$f \colon \overline{\Delta} \to \C$ is a continuous function,
holomorphic in $\Delta$, and that $\Gamma = \partial \Delta_1 \times \cdots \times \partial \Delta_n$
is oriented appropriately (each $\partial \Delta_k$ oriented positively).
Then for $z \in \Delta$
\begin{equation}
f(z) =
\frac{1}{{(2\pi i)}^n}
\int_{\Gamma}
\frac{f(\zeta_1,\zeta_2,\ldots,\zeta_n)}{(\zeta_1-z_1)(\zeta_2-z_2)\cdots(\zeta_n-z_n)}
\,
\d \zeta_1
\wedge
\d \zeta_2
\wedge
\cdots
\wedge
\d \zeta_n .
\end{equation}
\end{mtheorem}
Here, we stated a general result where $f$ is only continuous on $\bar{\Delta}$ and holomorphic on $\Delta$. We are going to cheat and use the short-hand notation:
\be
\dfrac{1}{\zeta-z}
\overset{\text{def}}{=}
\dfrac{1}{(\zeta_1-z_1)(\zeta_2-z_2)\cdots(\zeta_n-z_n)}\,,
\ee
together with
\be
\d \zeta
\overset{\text{def}}{=}
\d \zeta_1
\wedge
\d \zeta_2
\wedge
\cdots
\wedge
\d \zeta_n\, .
\ee
This allows us to write the Cauchy integral formula in a concise way:
\be
f(z) =
\frac{1}{{(2\pi i)}^n}
\int_{\Gamma}
\frac{f(\zeta)}{\zeta-z}
\,
\d\zeta .
\ee
The Cauchy integral formula shows an important and subtle point about the holomorphic functions in several variables: The function $f(z)$ for $z \in \Delta$ is determined in terms of its values on the set $\Gamma$, which is much smaller than the boundary of the polydisc $\partial \Delta$. In fact, $\Gamma$ (a torus) is of real dimension $n$, while the boundary has real dimension $2n-1$. This is the first big difference compared to 1D. We call $\Gamma$ the \emph{distinguished boundary}. Let us draw the unit bidisc:
\be
\begin{adjustbox}{valign=c}
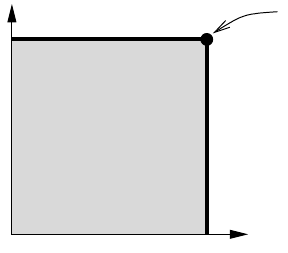
\end{adjustbox}
\ee
In this case, the set $\Gamma$ is a $2$-dimensional torus, like the surface of a donut. On the other hand, the set $\partial \D^2 = (\partial \D \times \overline{\D}) \cup ( \overline{\D} \times \partial \D)$ is the union of two filled donuts, or more precisely, it is both the inside and the outside of the donut put together, and these two things meet on the surface of the donut. So, the set $\Gamma$ is quite small in comparison to the entire bounary $\partial \D^2$. 

\subsection{Power series representation}

Writing expressions out in all the components can be a pain.
For $\alpha \in \N_{0}^{n}$ (where $\N_0 = \N \cup \{0\}$), we will cheat some more and use the multi-index notation to deal with the more complicated formulas:
\be
z^\alpha  \overset{\text{def}}{=} z_1^{\alpha_1}z_2^{\alpha_2} \cdots
z_n^{\alpha_n} ,
\qquad
\frac{\partial^{\sabs{\alpha}}}{\partial z^\alpha} \overset{\text{def}}{=}
\frac{\partial^{\alpha_1}}{\partial z_1^{\alpha_1}}
\frac{\partial^{\alpha_2}}{\partial z_2^{\alpha_2}}
\cdots
\frac{\partial^{\alpha_n}}{\partial z_n^{\alpha_n}} ,
\ee
\be
\alpha!  \overset{\text{def}}{=} \alpha_1!\alpha_2! \cdots \alpha_n! ,
\qquad
\alpha+1  \overset{\text{def}}{=} (\alpha_1+1,\alpha_2+1, \cdots \alpha_n+1) .
\ee

Let $\Delta$ be a polydisc with distinguished boundary $\Gamma$,
centered at $a$, of polyradius $\rho$.
Suppose $f$ is continuous on $\overline{\Delta}$, holomorphic on $\Delta$.
We will now differentiate under the integral in the Cauchy integral formula. This implies that $f$ is infinitely $\C$-differentiable and
\be
\frac{\partial^{\sabs{\alpha}}f}{\partial z^\alpha} (z) =
\frac{1}{{(2\pi i)}^n}
\int_{\Gamma}
\frac{\alpha! \, f(\zeta)}{{(\zeta-z)}^{\alpha+1}}
\,
\d \zeta .
\ee
From this, we get
\emph{the Cauchy estimates}, which are bounds on the growth of derivatives of $f$:
\begin{equation}
\abs{
\frac{\partial^{\sabs{\alpha}}f}{\partial z^\alpha} (a)
}
\leq \frac{\alpha! \, \snorm{f}_\Gamma}{\rho^\alpha} 
=
\frac{\alpha! \, \sup_{z\in \Gamma} \sabs{f(z)}}{\rho^\alpha} .
\end{equation}
In particular, the coefficients of the power series depend only on the derivatives of $f$ at $a$ and not the specific polydisc $\Delta$ used above.

\begin{mdcorollary}
The ``correct'' topology on $\sO(U)$
is uniform convergence on compacts
(normal convergence).  If $f_n \in \sO(U)$ and $f_n \to f$ uniformly
on compacts, then $f \in \sO(U)$ and $f_n^{(\ell)} \to f^{(\ell)}$ uniformly
on compacts.
\end{mdcorollary}

As in one variable, we can introduce the geometric series in several variables. For $z \in \D^n$ (unit polydisc):
\be
\frac{1}{1-z} =
\frac{1}{(1-z_1)(1-z_2)\cdots(1-z_n)} =
\left(
\sum_{k=0}^\infty {z_1}^k
\right)
\left(
\sum_{k=0}^\infty {z_2}^k
\right)
\cdots
\left(
\sum_{k=0}^\infty {z_n}^k
\right)
\ee
\be
=
\sum_{\alpha_1=0}^\infty
\sum_{\alpha_2=0}^\infty
\cdots
\sum_{\alpha_n=0}^\infty
\left(
{z_1}^{\alpha_1}
{z_n}^{\alpha_2}
\cdots
{z_n}^{\alpha_n}
\right)
=
\sum_{\alpha} z^\alpha .
\ee
Power series 
$\sum_{\alpha} c_\alpha (z-a)^\alpha$ converges
absolutely uniformly on compact subsets of its
\emph{domain of convergence}
(interior of the set where it converges).

\begin{mdexample}
In $\C^2$, the power series
\be
\sum_{k=0}^\infty z_1 z_2^k\,,
\ee
converges absolutely on the set
\be
\bigl\{ z \in \C^2 : \sabs{z_2} < 1 \bigr\}
\cup
\bigl\{ z \in \C^2 : z_1 = 0 \bigr\}\,,
\ee
and nowhere else. This set is not a polydisc.
It is neither an open set nor a closed set. Its closure is not the closure of the domain of convergence, which is the set $\bigl\{ z \in \C^2 : \sabs{z_2} < 1 \bigr\}$.
\end{mdexample}
\begin{mdexample}
The power series
\be
\sum_{k=0}^\infty z_1^k z_2^k\,,
\ee
converges absolutely exactly on the set
\be
\bigl\{ z \in \C^2 : \sabs{z_1 z_2} < 1 \bigr\}.
\ee
The picture of this domain is more complicated than that of a polydisc:
\be
\begin{adjustbox}{valign=c}
\scalebox{1}{
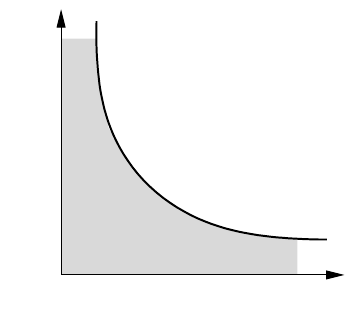
}
\end{adjustbox}
\ee
\end{mdexample}

Let $\Delta = \Delta_\rho(a) \subset \C^n$ be a polydisc
and $f$ is holomorphic in a neighborhood of $\overline{\Delta}$,
let $\Gamma$ be the distinguished boundary of $\Delta$.
In the Cauchy integral formula,
\be
f(z)
=
\frac{1}{{(2\pi i)}^n}
\int_{\Gamma}
\frac{f(\zeta)}{\zeta-z}
\,
\d \zeta \, ,
\ee
let us expand the Cauchy kernel as
\be
\frac{1}{\zeta-z} =
\frac{1}{\zeta-a}
\left(
\frac{1}{1-\frac{z-a}{\zeta-a}}
\right)
=
\frac{1}{\zeta-a}
\sum_{\alpha}
{\left(\frac{z-a}{\zeta-a}\right)}^\alpha \, .
\ee
Here, we make sure to interpret the formulas correctly as $\frac{1}{\zeta - z} = \frac{1}{(\zeta_1 - z_1)\cdots (\zeta_n - z_n)}$ and so on.
The multivariable geometric series is a product of geometric series in one variable, and geometric series in one variable are uniformly absolutely convergent on compact subsets of the unit disc. This allows us to prove the following theorem.
\begin{mtheorem}
For $z \in \Delta$,
\be
\hspace{-0.3cm}
f(z) =
\sum_{\alpha}
c_{\alpha}
{(z-a)}^{\alpha} ,
\quad \text{where} \quad
c_\alpha
=
\frac{1}{\alpha!} \frac{\partial^{\sabs{\alpha}}f}{\partial
z^\alpha} (a)
=
\frac{1}{{(2\pi i)}^n}
\int_{\Gamma}
\frac{f(\zeta)}{{(\zeta-a)}^{\alpha+1}}
\,
\d \zeta .
\ee
Conversely, if $f$ is defined by a power series, then it is holomorphic.
\end{mtheorem}
The proof of the first statement is a simple computation and application of the Fubini theorem or uniform convergence just as in one variable. The converse statement can be proven in different ways. For example, it follows by applying the Cauchy--Riemann equations to the series termwise.

This is in fact the first place where the theory of several complex variables becomes annoying.
\begin{mtheorem}[Identity]
Let $U \subset \C^n$ be a domain (connected open set) and let
$f \colon U \to \C$ be holomorphic.
If $f|_N \equiv 0$ for a nonempty open subset $N \subset U$,
then $f \equiv 0$.
\end{mtheorem}
Here we encounter a difference from the 1D cases: The zero set of a holomorphic function
in 2 or more complex variables is always large (it always has limit points). The above theorem is often used to show that if two holomorphic functions $f$ and $g$ are equal on a small open set, then $f \equiv g$.

\begin{mtheorem}[Maximum principle]
Let $U \subset \C^n$ be a domain.
Let $f \colon U \to \C$ be holomorphic and suppose that $\sabs{f(z)}$
attains a local maximum at some $a \in U$.  Then $f \equiv f(a)$.
\end{mtheorem}
Here, the argument goes back to 1D. The proof involves using the maximum principle on every 1D subspace.

\subsection{Inequivalence of ball and polydisc}

We say that $f \colon U \to V$ is a \emph{biholomorphism}
(and $U$ and $V$ are \emph{biholomorphic}) if $f$ is
bijective, holomorphic, and $f^{-1}$ is holomorphic. One of the main questions in complex analysis is to classify domains up to biholomorphic transformations. 

In one variable, there is the rather striking theorem due to Riemann: If $U \subset \C$ is a nonempty simply connected domain such that $U \neq \C$, then $U$ is biholomorphic to $\D$. So essentially, in 1D a topological property on $U$ is enough to classify a whole class of domains. It is one of the reasons why studying the disc is so important in one variable and why many theorems are stated for the disc only.
There is no Riemann Mapping Theorem in several dimensions! In fact, it is not difficult to find 2D domains that are not biholomorphic.
\begin{mtheorem}[Poincar\'e, 1907]
$\bB_2$ and $\D^2$ are not biholomorphic.
\end{mtheorem}
The first complete proof was given by Henri Cartan in 1931, though popularly the theorem is attributed to Poincar\'e.
Note that both domains are simply connected (have no holes) and they are the two most obvious generalizations of the disc to two variables. They are homeomorphic, that is, topology does not see any difference.

We need to introduce some constructions before attempting a proof of the above theorem. A nonconstant holomorphic mapping $\varphi \colon \D \to \C^n$ is called an analytic disc.
It plays an important role in the geometry of several complex variables. Essentially, it allows us to apply one-variable results in several variables. It is especially important in understanding the boundary behavior of holomorphic functions and features prominently in complex geometry. Often, we call the image $\Delta = \varphi(\D)$ the analytic disc rather than the mapping.

\begin{mdproposition}
The unit sphere $S^{2n-1} = \partial \bB_n \subset \C^n$
contains no analytic discs.\\

\noindent\emph{Proof:} 
Suppose there is a holomorphic function $g \colon \D \to S^{2n-1} \subset \C^n$:
\be
\sabs{g_1(z)}^2 + \sabs{g_2(z)}^2 + \cdots + \sabs{g_n(z)}^2 = 1\,, 
\ee
for all $z \in \D$.
Without loss of generality (after composing with a unitary matrix) suppose that $g(0)=(1,0,\cdots,0)$. Notice that $g_1(0)=1$ and $|g_1(z)| \leq 1$.  The maximum principle says that $g_1$ attains its maximum for all $z \in \D$ which implies that $g_j(z)=0$ for all $j=2,\ldots,n$ on all $z \in \D$. Hence $g$ has to be a constant function and not an analytic disc.  \hfill$\square$
\end{mdproposition}
The fact that the sphere contains no analytic discs is the most important geometric distinction between the boundary of the polydisc and the sphere.

Let us now sketch the proof of the Poincar\'e theorem using pictures. We will use proof by contradiction. To this end, suppose $f: \D^2 \to \B_2$ is a biholomorphism.
Let us pick a disc for fixed $z_1 = \zeta$ and a sequence $w_k \to \e^{i\theta}$. It looks like this:
\be
\begin{adjustbox}{valign=c}
\scalebox{1}{
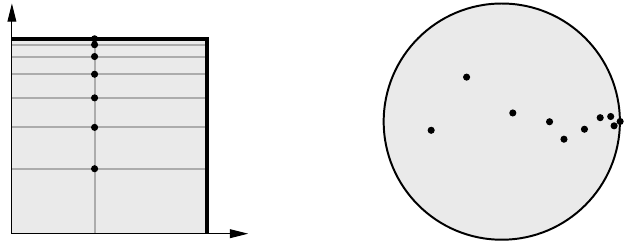
}
\end{adjustbox}
\ee
The idea is to show that (after passing to a subsequence via Montel) $\zeta \mapsto f(\zeta,w_k)$ converges to a holomorphic map to the sphere. Therefore, it has to be a constant.
More precisely, the derivative of $\zeta \mapsto f(\zeta,w_k)$ goes to zero
for every $\e^{i\theta}$
and every $\{ w_k \}$. This implies that $\frac{\partial f}{\partial z_1} \equiv 0$ and by symmetry also
$\frac{\partial f}{\partial z_2} \equiv 0$. Therefore $f$ has to be a constant and we run into a contradiction.

The proof says that the reason why there is not even a proper mapping is the fact that the boundary of the polydisc contains analytic discs, whereas the sphere does not. The proof extends easily to higher dimensions, as well. The key takeaway is that in several complex variables, the geometry of the boundary makes a difference if one wants to determine if two domains are equivalent. The domain topology is not enough.

\subsection{Cartan's uniqueness theorem}

Where did Schwarz's lemma go? The following theorem is its analogue for several variables.
\begin{mtheorem}[Cartan]
Suppose $U \subset \C^n$ is a \textbf{bounded} domain, $a \in U$,
$f \colon U \to U$ is a holomorphic mapping, $f(a) = a$,
and $Df(a)$ is the identity.  Then $f(z) = z\,$ for all $z \in U$.
\end{mtheorem}
Here, we bolded the word ``bounded'' which is crucial. Counterexamples can be found if $U$ is unbounded.

The argument is to use Cauchy estimates on the first
nonzero nonlinear term of the series of $f^\ell = f \circ f \circ \cdots
\circ f$. The result can be used to compute automorphism groups just as in 1D.
Every automorphism of $\D^n$ is of the form
\be
z\mapsto P \left(
\e^{i\theta_1} \frac{a_1-z_1}{1-\bar{a}_1z_1} ,
\e^{i\theta_2} \frac{a_2-z_2}{1-\bar{a}_2z_2} , \ldots,
\e^{i\theta_n} \frac{a_n-z_n}{1-\bar{a}_nz_n} \right),
\ee
where $\theta \in \R^n$, $a \in \D^n$, and $P$ is a permutation matrix.
On the other hand, every automorphism of $\bB_n$ is of the form
\be
z \mapsto U \frac{a-P_az-s_a(I-P_a)z}{1-\linnprod{z}{a}}\, ,
\ee
where
$a \in \bB_n$, $U$ is a unitary matrix, 
$s_a = \sqrt{1-\snorm{a}^2}$, and
$P_a z = \frac{\linnprod{z}{a}}{\linnprod{a}{a}}a$.

\subsection{Riemann extension theorem, zero sets, and injective maps}

In 1D, if a function is holomorphic in $U \setminus \{p\}$ and locally bounded in $U$ (for every $q \in U$ there is a neighborhood $W$ of $q$ such that $f$ is bounded on $W \cap (U \setminus \{p\})$), then the function extends holomorphically to $U$. In several variables, the same theorem holds, and the analogue of a single point is the zero set of a holomorphic function.

\begin{mtheorem}[Riemann extension theorem]
Let $U \subset \C^n$ be a domain,  $g \in \sO(U)$, $g \not\equiv
0$, and $N = g^{-1}(0)$.
If $f \in \sO(U \setminus N)$ is locally bounded in $U$,
then there exists a unique $F \in \sO(U)$ such that $F|_{U \setminus N} = f$.
\end{mtheorem}
Its proof involves cutting $N$ ``transversely'' by complex lines, applying the 1D
Riemann mapping theorem and using the Cauchy formula as glue.

The set of zeros of a holomorphic function has a nice structure at most points, as codified in the following theorem.
\begin{mtheorem}
Let $U \subset \C^n$ be a domain,
$f \in \sO(U)$, $f \not\equiv 0$, and $N = f^{-1}(0)$.
Then there exists an open and dense
$N_{\text{reg}} \subset N$
such that at each $p \in N_{\text{reg}}$, after possibly reordering variables,
$N$ can be locally written as
\begin{equation}
z_n = g(z_1,\ldots,z_{n-1})\,,
\end{equation}
for a holomorphic function $g$.
\end{mtheorem}
The proof is to consider all possible derivatives of $f$. One of them
must be nonzero somewhere on $N$ (by analyticity). Then one applies the implicit
function theorem.

For holomorphic $f \colon U \subset \C^n \to \C^n$, let us write the holomorphic Jacobian as
\be
Df = \left[ \frac{\partial f_k}{\partial z_\ell} \right]_{k\ell}.
\ee
Note that 
$\sabs{\det Df}^2 = \det D_\R f$, where $D_\R f$ is
the real Jacobian matrix.

\begin{mtheorem}
Suppose $U \subset \C^n$ is open and $f \colon U \to \C^n$ is
holomorphic and one-to-one.
Then $\det Df$ is never zero on $U$.
\end{mtheorem}
In 1D, every holomorphic function $f$ can, in the proper local holomorphic coordinates, be written as $z^d$ for $d=0,1,2,\ldots$, up to a constant. Such a simple result does not hold in several complex variables in general, but if the mapping is locally one-to-one, the present theorem says that such a mapping can be locally written as the identity.
The proof of this theorem essentially reduces to the 1D statement, but not trivially.

Therefore, if a holomorphic map $f \colon U \to V$ is bijective
for two open sets $U,V \subset \C^n$, then $f$ is biholomorphic.
\begin{mdexample}
The theorem does not hold in different dimensions.
$f \colon \C \to \C^2$ given by $z \mapsto (z^2,z^3)$ is one-to-one
and onto the cusp ($z_1^3-z_2^2=0$), but $f'(0) = 0$.
\end{mdexample}

\section[Convexity and pseudoconvexity]{\label{sec:Sean-lecture-2}Convexity and pseudoconvexity\\
\normalfont{\textit{Sean Curry}}}

\subsection{Pseudoconvexity}

The motivating question of this lecture is: Which domains can be domains of maximal analytic continuation of a holomorphic function?

A central concept which will be introduced is \emph{pseudoconvexity}. It will answer the above question. In fact, we will distinguish between three different notions:
\begin{itemize}
    \item[$\diamond$] Convexity (linear convexity in the usual sense),
    \item[$\diamond$] $\C$-linear convexity,
    \item[$\diamond$] Pseudoconvexity.\footnote{In a tangential direction, Sean started by mentioning some places where the topics of Cauchy-Riemann geometry and (strong) pseudoconvexity arise in connection with physics (``tasty bits'' that unfortunately go beyond the scope of these introductory lectures). The work of Fefferman \cite{Fefferman1974,Fefferman1976,Fefferman1979} in the 1970s on strongly pseudoconvex domains inspired the work Fefferman and Graham \cite{FeffermanGraham1985} (cf.\ \cite{FeffermanGraham2012}) on conformal invariants, where they developed the boundary expansions for the asymptotically AdS metrics that are used in the AdS/CFT correspondence; Fefferman's result \cite{Fefferman1974} also showed that the biholomorphically invariant geometry of a bounded strongly pseudoconvex domain in $\mathbb{C}^n$ is equivalent to the Cauchy-Riemann geometry of the boundary, a prototypical example of a ``bulk-boundary correspondence'' with strong connections to AdS/CFT. Cauchy-Riemann geometry also arises in various ways in twistor theory and in the antecedent work of Robinson, Trautman and others, where Cauchy-Riemann structures arise from shear-free null geodesic congruences in spacetimes (which are ``twisting'' if the Cauchy-Riemann structure is strongly pseudoconvex). In this setting a connection was found between Maxwell's equations and the tangential Cauchy-Riemann equations and then between Einstein's equations and the tangential Cauchy-Riemann equations. This played an important role in the discovery and study of algebraically special spacetimes such as the Kerr black hole solution \cite{RobinsonTrautman1961,Kerr1963,PenroseRindler-II,HillLewandowskiNurowski2008}.
    }
\end{itemize}
For example, a bean shape is not convex but it may be pseudoconvex. We will elaborate on these notions in what follows.

In this talk, we will focus on domains rather than functions.
It turns out that not every domain in $\C^n$ is a natural domain for holomorphic functions.\footnote{Sean is using the notation $\subseteq$ to mean the same as Jiří's $\subset$. Both symbols denote a subset that does not need to be proper.}
\begin{mddefinition}
    Let $U \subseteq \C^n$ be a domain (a connected open set). The set $U$ is called a \emph{domain of holomorphy} if there do not exist nonempty open sets $V$ and $W$ of $\C^n$ with $V \subseteq U \cap W$, $W \not\subseteq U$, and $W$ connected, such that for every function $f \in \sO(U)$ there exists an $F \in \sO(W)$ with $F|_{V} = f|_{V}$.
\end{mddefinition}
The idea here is that if a domain $U$ is not a domain of holomorphy and both $V$, $W$ exist as in the definition above, then $f$ ``extends across the boundary'' somewhere. This is illustrated with the following picture, where $V$ acts as ``glue'' between the other two domains:
\be
\begin{adjustbox}{valign=c}
\scalebox{1}{
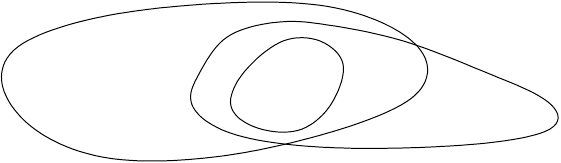
}
\end{adjustbox}
\ee
Note that we also can have non-trivial monodromy as follows:
\begin{equation}
    \adjustbox{valign=c}{\tikzset{every picture/.style={line width=0.5pt}}
\begin{tikzpicture}[x=0.75pt,y=0.75pt,yscale=-1,xscale=1]

\draw  [draw opacity=1][fill=none, fill opacity=0.2 ] (428.74,139.35) .. controls (402.84,135.48) and (414.14,110.84) .. (433.63,112.51) .. controls (453.11,114.18) and (471.16,114.41) .. (483.75,117.18) .. controls (496.34,119.94) and (563.45,152.9) .. (497,167.33) .. controls (430.55,181.77) and (346,157.33) .. (440,153.33) .. controls (534,149.33) and (454.63,143.22) .. (428.74,139.35) -- cycle ;
\begin{scope}[xshift=190pt,yshift=20pt]
\draw [color=Maroon  ,draw opacity=1 ]   (200,120.5) .. controls (201.67,122.17) and (201.67,123.83) .. (200,125.5) .. controls (198.33,127.17) and (198.33,128.83) .. (200,130.5) .. controls (201.67,132.17) and (201.67,133.83) .. (200,135.5) .. controls (198.33,137.17) and (198.33,138.83) .. (200,140.5) .. controls (201.67,142.17) and (201.67,143.83) .. (200,145.5) .. controls (198.33,147.17) and (198.33,148.83) .. (200,150.5) .. controls (201.67,152.17) and (201.67,153.83) .. (200,155.5) .. controls (198.33,157.17) and (198.33,158.83) .. (200,160.5) .. controls (201.67,162.17) and (201.67,163.83) .. (200,165.5) -- (200,167) -- (200,167) ;
\draw  [draw opacity=0][fill=Maroon  ,fill opacity=1 ] (202.5,120.5) .. controls (202.5,119.12) and (201.38,118) .. (200,118) .. controls (198.62,118) and (197.5,119.12) .. (197.5,120.5) .. controls (197.5,121.88) and (198.62,123) .. (200,123) .. controls (201.38,123) and (202.5,121.88) .. (202.5,120.5) -- cycle ;
\end{scope}
\draw  [draw opacity=1 ][fill=none ,fill opacity=0.2 ] (363.67,101) .. controls (383.67,91) and (375.67,81) .. (416.67,82) .. controls (457.67,83) and (468.67,112) .. (445.67,146) .. controls (422.67,180) and (377.67,191) .. (357.67,161) .. controls (337.67,131) and (343.67,111) .. (363.67,101) -- cycle ;
\draw  [draw opacity=1 ][fill=white  ,fill opacity=1 ] (422.13,126.01) .. controls (422.13,119.66) and (427.27,114.51) .. (433.63,114.51) .. controls (439.98,114.51) and (445.13,119.66) .. (445.13,126.01) .. controls (445.13,132.36) and (439.98,137.51) .. (433.63,137.51) .. controls (427.27,137.51) and (422.13,132.36) .. (422.13,126.01) -- cycle ;
\draw  [draw opacity=1][fill=none, fill opacity=0.2 ] (422.13,126.01) .. controls (422.13,119.66) and (427.27,114.51) .. (433.63,114.51) .. controls (439.98,114.51) and (445.13,119.66) .. (445.13,126.01) .. controls (445.13,132.36) and (439.98,137.51) .. (433.63,137.51) .. controls (427.27,137.51) and (422.13,132.36) .. (422.13,126.01) -- cycle ;
\draw (425.67,118.4) node [anchor=north west][inner sep=0.75pt]    {$V$};
\draw (493.67,135.4) node [anchor=north west][inner sep=0.75pt]    {$W$};
\draw (351.67,116.4) node [anchor=north west][inner sep=0.75pt]    {$U$};
\end{tikzpicture}}
\end{equation}
Here in principle the domain $W$ can wrap around a branch cut, in which case the function ``locally extends'' around it. A function $F$ defined on $W$ is said to extend the function $f$ on $U$ if $F|_V=f|_V$ (for some open set $V\subseteq U\cap W$); the functions $F$ and $f$ may differ on other parts of $U\cap W$, i.e.\ the extension may give rise to a multivalued function. For example, the principal branch of the logarithm function on the cut plane is extendable across the branch cut away from the origin.

\begin{mdremark}
One can show (see, e.g., \cite[Thm.~2.5.5]{hormander1973introduction}) that if $\Omega$ is a domain of holomorphy, then there exists a function $f \in \sO(\Omega)$ that cannot be analytically continued past $\Omega$. However, any individual $f \in \sO(\Omega)$ might admit an analytic continuation outside of $\Omega$.
\end{mdremark}

\begin{mdexample}\label{ex:disc-is-dom-hol}
    The unit disc $\D$ is a domain of holomorphy since we can always construct functions with singularities on any point of $\partial \D$. For example, if $\{1\} \in W$, then $\frac{1}{1-z}$ does not extend beyond $z=1$.
    Similarly, if $\{\e^{i\theta}\} \in W$, then $\frac{1}{\e^{i\theta} - z}$ does not extend beyond $z=\e^{i\theta}$.
\end{mdexample}
\begin{mdexample}
    Alternatively, one can ask what functions have $\D$ as the maximal domain? We can construct such functions, with a dense set of singularities on the boundary $\partial \D$, by defining
    \begin{equation}
        f(z) = \sum_{k=0}^{\infty} z^{k!} \qquad\text{or}\qquad f(z) = \sum_{k=0}^{\infty} z^{a^k}\,,
    \end{equation}
    with $a \in \Z, a \geq 2$. Consider the first one.
    By plugging in $z = \e^{2 \pi i \frac{p}{q}}$ for any fraction $\frac{p}{q}$ we get that for any $k \geq q$ in the sum, $\big(\e^{2 \pi i \frac{p}{q}} \big)^{k!}=1$, so the series defining $f(z)$ diverges at any $z=\e^{2 \pi i \frac{p}{q}}$ and from the sparsity of the coefficients we can conclude that $f(z)$ tends to infinity as $z$ tends to any such boundary point. These points form a dense set of singularities on $\partial \D$, so that $f(z)$ cannot be analytically continued across any boundary point. Functions with such an obstruction to analytic continuations are called \emph{lacunary} functions.
\end{mdexample}

\begin{mdremark}
    The argument from Ex.~\ref{ex:disc-is-dom-hol} showing that $\D$ is a domain of holomorphy can be applied to any domain $U \subseteq \C$. Thus, any domain $U \subseteq \C$ is a domain of holomorphy, since we can place poles at any point on the boundary of $U$.
\end{mdremark}

Now we want to understand how domains of holomorphy can look like in several complex variables. Let us first consider an example.
\begin{mdexample}
The unit ball $\B^n \subseteq \C^n$ is a domain of holomorphy. \\

\emph{Proof:} If $(1,0,\ldots,0) \in W$, then the function $f(z) = \frac{1}{1-z_1}$ does not extend across $\B^n$. 
\be
\adjustbox{valign=c}{\tikzset{every picture/.style={line width=0.75pt}} 
\begin{tikzpicture}[x=0.75pt,y=0.75pt,yscale=-1,xscale=1]
\draw   (191,154.5) .. controls (191,128.82) and (211.82,108) .. (237.5,108) .. controls (263.18,108) and (284,128.82) .. (284,154.5) .. controls (284,180.18) and (263.18,201) .. (237.5,201) .. controls (211.82,201) and (191,180.18) .. (191,154.5) -- cycle ;
\draw    (284,104) -- (284,205) ;
\draw  [draw opacity=0][fill={rgb, 255:red, 0; green, 0; blue, 0 }  ,fill opacity=1 ] (281.38,157.13) .. controls (281.38,155.68) and (282.55,154.5) .. (284,154.5) .. controls (285.45,154.5) and (286.63,155.68) .. (286.63,157.13) .. controls (286.63,158.57) and (285.45,159.75) .. (284,159.75) .. controls (282.55,159.75) and (281.38,158.57) .. (281.38,157.13) -- cycle ;
\draw  [draw opacity=0] (284,157.13) .. controls (270.05,159.44) and (254.06,160.76) .. (237.06,160.77) .. controls (220.76,160.77) and (205.38,159.58) .. (191.84,157.45) -- (237.05,130.77) -- cycle ; \draw   (284,157.13) .. controls (270.05,159.44) and (254.06,160.76) .. (237.06,160.77) .. controls (220.76,160.77) and (205.38,159.58) .. (191.84,157.45) ;  
\draw  [draw opacity=0][dash pattern={on 4.5pt off 4.5pt}] (191.84,157.4) .. controls (205.79,155.1) and (221.78,153.79) .. (238.78,153.79) .. controls (255.08,153.79) and (270.46,154.99) .. (284,157.13) -- (238.78,183.79) -- cycle ; \draw  [dash pattern={on 4.5pt off 4.5pt}] (191.84,157.4) .. controls (205.79,155.1) and (221.78,153.79) .. (238.78,153.79) .. controls (255.08,153.79) and (270.46,154.99) .. (284,157.13) ;  
\draw[<-,color=Gray]    (291,188) -- (329,167.25) node[pos=1,right,scale=0.9]{$z_1=1$};
\end{tikzpicture}}
\ee
By symmetry, other points can be handled by composing the function $f(z)$ with a rotation, giving a function of the form $\frac{1}{\text{(affine linear)}}$ with a pole on a complex affine subspace of $\mathbb{C}^n$ that passes through the chosen point on $\partial \mathbb{B}^n$ (and is tangent to $\partial \mathbb{B}^n$ at that point). 
\be
\adjustbox{valign=c}{\tikzset{every picture/.style={line width=0.75pt}}
\begin{tikzpicture}[x=0.75pt,y=0.75pt,yscale=-1,xscale=1]
\draw   (191,154.5) .. controls (191,128.82) and (211.82,108) .. (237.5,108) .. controls (263.18,108) and (284,128.82) .. (284,154.5) .. controls (284,180.18) and (263.18,201) .. (237.5,201) .. controls (211.82,201) and (191,180.18) .. (191,154.5) -- cycle ;
\draw  [draw opacity=0] (284,157.13) .. controls (270.05,159.44) and (254.06,160.76) .. (237.06,160.77) .. controls (220.76,160.77) and (205.38,159.58) .. (191.84,157.45) -- (237.05,130.77) -- cycle ; \draw   (284,157.13) .. controls (270.05,159.44) and (254.06,160.76) .. (237.06,160.77) .. controls (220.76,160.77) and (205.38,159.58) .. (191.84,157.45) ;  
\draw  [draw opacity=0][dash pattern={on 4.5pt off 4.5pt}] (191.84,157.4) .. controls (205.79,155.1) and (221.78,153.79) .. (238.78,153.79) .. controls (255.08,153.79) and (270.46,154.99) .. (284,157.13) -- (238.78,183.79) -- cycle ; \draw  [dash pattern={on 4.5pt off 4.5pt}] (191.84,157.4) .. controls (205.79,155.1) and (221.78,153.79) .. (238.78,153.79) .. controls (255.08,153.79) and (270.46,154.99) .. (284,157.13) ;  
\draw  [color=RoyalBlue ,draw opacity=1 ][fill=RoyalBlue ,fill opacity=0.18 ] (255.72,100.46) -- (326.07,166.34) -- (256.28,155.79) -- (185.93,89.91) -- cycle ;
\draw [color=Maroon  ,draw opacity=1 ]   (220.82,95.19) -- (291.18,161.06) ;
\draw  [draw opacity=0][fill={rgb, 255:red, 0; green, 0; blue, 0 }  ,fill opacity=1 ] (253.38,128.13) .. controls (253.38,126.68) and (254.55,125.5) .. (256,125.5) .. controls (257.45,125.5) and (258.63,126.68) .. (258.63,128.13) .. controls (258.63,129.57) and (257.45,130.75) .. (256,130.75) .. controls (254.55,130.75) and (253.38,129.57) .. (253.38,128.13) -- cycle ;
\draw[<-,color=Maroon]    (269,135) -- (303,106.25) node[pos=1,right,scale=0.9]{Affine linear = 0} node[pos=1,right,yshift=-15,scale=0.9]{sits inside \textcolor{RoyalBlue}{$T_{p} S^{2n-1}$}};
\end{tikzpicture}}
\ee
\end{mdexample}
\begin{mdremark}
    The same construction clearly apples to any convex domain (recall that, for simplicity, we are restricting to domains with smooth boundary). In particular, any convex (or $\C$-linearly convex) domain $U \subseteq \C^n$ is a domain of holomorphy. (Here, $\C$-linearly convex means convex in complex tangential directions.)

\end{mdremark}
The key point is that, in order to have a domain of holomorphy in $\C^n$, we need room to ``fit'' the singularities which now lie on hypervarieties. Note that the ball $\B^n$ is convex and is in fact a domain of holomorphy. As noted in the preceding remark, this observation works more generally.
\begin{mtheorem}
Any convex domain $U \subseteq \C^n$ is a domain of holomorphy.
\end{mtheorem}
The question now becomes whether convexity is actually needed.

\subsection{Non-convex domain of holomorphy}

Domains of holomorphy are often not geometrically convex, so classical convexity is not the correct notion but it is in the right direction. Before giving an example of a non-convex domain of holomorphy, we show that by a biholomorphic change of coordinates the unit ball in $\mathbb{C}^n$ (which is strongly convex) can be realized as an unbounded domain that is convex but not strongly convex.

\be
\label{cayley-figure}
\adjustbox{valign=c,scale={0.73}{0.73}}{\tikzset{every picture/.style={line width=0.75pt}}    
\begin{tikzpicture}[x=0.75pt,y=0.75pt,yscale=-1,xscale=1]
\draw  [draw opacity=0] (647.01,85.8) .. controls (643,114.9) and (625.53,144.5) .. (607.96,151.95) .. controls (607.39,152.19) and (606.82,152.41) .. (606.25,152.6) -- (615.18,99.19) -- cycle ; \draw   (647.01,85.8) .. controls (643,114.9) and (625.53,144.5) .. (607.96,151.95) .. controls (607.39,152.19) and (606.82,152.41) .. (606.25,152.6) ;  
\draw  [draw opacity=0] (606.25,152.6) .. controls (603.45,153.41) and (600.59,153.54) .. (597.72,152.9) .. controls (585.33,150.13) and (576.66,133.72) .. (575.17,113) -- (604.94,100.15) -- cycle ; \draw   (606.25,152.6) .. controls (603.45,153.41) and (600.59,153.54) .. (597.72,152.9) .. controls (585.33,150.13) and (576.66,133.72) .. (575.17,113) ;  
\draw    (417.17,112) -- (575.17,113) ;
\draw    (442,152.5) -- (602,153.5) ;
\draw  [draw opacity=0] (448.25,151.6) .. controls (445.45,152.41) and (442.59,152.54) .. (439.72,151.9) .. controls (427.33,149.13) and (418.66,132.72) .. (417.17,112) -- (446.94,99.15) -- cycle ; \draw   (448.25,151.6) .. controls (445.45,152.41) and (442.59,152.54) .. (439.72,151.9) .. controls (427.33,149.13) and (418.66,132.72) .. (417.17,112) ;  
\draw    (489.01,84.8) -- (647.01,85.8) ;
\draw  [draw opacity=0] (489.01,84.8) .. controls (487.71,94.29) and (484.96,103.84) .. (481.25,112.66) -- (457.18,98.19) -- cycle ; \draw   (489.01,84.8) .. controls (487.71,94.29) and (484.96,103.84) .. (481.25,112.66) ;  
\draw  [draw opacity=0][dash pattern={on 4.5pt off 4.5pt}] (481.25,112.66) .. controls (473.59,130.93) and (461.77,146.07) .. (449.9,151.1) .. controls (449.32,151.35) and (448.75,151.57) .. (448.19,151.76) -- (457.11,98.35) -- cycle ; \draw  [dash pattern={on 4.5pt off 4.5pt}] (481.25,112.66) .. controls (473.59,130.93) and (461.77,146.07) .. (449.9,151.1) .. controls (449.32,151.35) and (448.75,151.57) .. (448.19,151.76) ;  
\draw[<-]    (526,44.67) -- (526,113.5) node[pos=0,right]{$\text{Im}(\omega)$};
\draw  [dash pattern={on 0.84pt off 2.51pt}]  (526,113.5) -- (526,136) ;
\draw  [dash pattern={on 0.84pt off 2.51pt}]  (526,136) -- (581,136.5) ;
\draw[->]    (581,136.5) -- (648,136.5) node[pos=1,below]{$\text{Re}(\omega)$};
\draw  [dash pattern={on 0.84pt off 2.51pt}]  (526,136) -- (509,151.5) ;
\draw[<-]    (466,184.5) -- (509,151.5) node[right,pos=0,xshift=3]{$z_{1}\ldots z_{n-1} \qquad (\text{Im}(\omega)>\|z\|^2)$};
\draw   (9,128.5) .. controls (9,102.82) and (29.82,82) .. (55.5,82) .. controls (81.18,82) and (102,102.82) .. (102,128.5) .. controls (102,154.18) and (81.18,175) .. (55.5,175) .. controls (29.82,175) and (9,154.18) .. (9,128.5) -- cycle ;
\draw  [color=RoyalBlue, draw opacity=0] (102,131.13) .. controls (88.05,133.44) and (72.06,134.76) .. (55.06,134.77) .. controls (38.76,134.77) and (23.38,133.58) .. (9.84,131.45) -- (55.05,104.77) -- cycle ; \draw   (102,131.13) .. controls (88.05,133.44) and (72.06,134.76) .. (55.06,134.77) .. controls (38.76,134.77) and (23.38,133.58) .. (9.84,131.45) ;  
\draw  [color=RoyalBlue, draw opacity=0][dash pattern={on 4.5pt off 4.5pt}] (9.84,131.4) .. controls (23.79,129.1) and (39.78,127.79) .. (56.78,127.79) .. controls (73.08,127.79) and (88.46,128.99) .. (102,131.12) -- (56.78,157.79) -- cycle ; \draw  [dash pattern={on 4.5pt off 4.5pt}] (9.84,131.4) .. controls (23.79,129.1) and (39.78,127.79) .. (56.78,127.79) .. controls (73.08,127.79) and (88.46,128.99) .. (102,131.12) ;  
\draw  [color=RoyalBlue, draw opacity=0] (142.45,100.66) .. controls (172.73,95.59) and (213.03,92.35) .. (257.29,92.04) .. controls (304.12,91.73) and (346.56,94.76) .. (377.41,99.93) -- (257.49,122.04) -- cycle ; \draw [->]   (142.45,100.66) .. controls (172.73,95.59) and (213.03,92.35) .. (257.29,92.04) .. controls (304.12,91.73) and (346.56,94.76) .. (377.41,99.93) ;  
\draw  [color=Maroon  ,draw opacity=1 ][fill=Maroon  ,fill opacity=0.11 ] (199.27,186) -- (358.02,186) -- (327.35,212.67) -- (168.6,212.67) -- cycle ;
\draw[color=Maroon, line width=1.25pt]   (236.75,199.33) .. controls (236.75,194.73) and (246.82,191) .. (259.25,191) .. controls (271.68,191) and (281.75,194.73) .. (281.75,199.33) .. controls (281.75,203.94) and (271.68,207.67) .. (259.25,207.67) .. controls (246.82,207.67) and (236.75,203.94) .. (236.75,199.33) -- cycle ;
\draw  [color=RoyalBlue  ,draw opacity=1 ][fill=RoyalBlue  ,fill opacity=0.17 ] (246.43,137.93) -- (319.34,216.99) -- (303.6,245.87) -- (230.7,166.81) -- cycle ;
\draw[color=RoyalBlue, line width=1.25pt]   (233,163.5) .. controls (269.59,201.05) and (274,204) .. (282.2,198.85) .. controls (286,192) and (285.59,184.05) .. (244.75,142);

\draw [draw opacity=1 ]   (307.55,158.71) -- (253.55,245.38) ;
\draw [draw opacity=1 ]   (200.25,161.33) -- (278,245.67) ;
\draw  [dash pattern={on 0.84pt off 2.51pt}]  (263,173.67) -- (263,240.67) ;
\draw  [color=Maroon, draw opacity=0][fill={rgb, 255:red, 0; green, 0; blue, 0 }  ,fill opacity=1 ] (260.38,193.67) .. controls (260.38,192.22) and (261.55,191.04) .. (263,191.04) .. controls (264.45,191.04) and (265.63,192.22) .. (265.63,193.67) .. controls (265.63,195.12) and (264.45,196.29) .. (263,196.29) .. controls (261.55,196.29) and (260.38,195.12) .. (260.38,193.67) -- cycle ;
\draw  [color=Maroon,draw opacity=0][fill={rgb, 255:red, 0; green, 0; blue, 0 }  ,fill opacity=1 ] (260.38,207.54) .. controls (260.38,206.09) and (261.55,204.92) .. (263,204.92) .. controls (264.45,204.92) and (265.63,206.09) .. (265.63,207.54) .. controls (265.63,208.99) and (264.45,210.17) .. (263,210.17) .. controls (261.55,210.17) and (260.38,208.99) .. (260.38,207.54) -- cycle ;
\draw  [color=Maroon, draw opacity=0][fill={rgb, 255:red, 0; green, 0; blue, 0 }  ,fill opacity=1 ] (260.38,230.67) .. controls (260.38,229.22) and (261.55,228.04) .. (263,228.04) .. controls (264.45,228.04) and (265.63,229.22) .. (265.63,230.67) .. controls (265.63,232.12) and (264.45,233.29) .. (263,233.29) .. controls (261.55,233.29) and (260.38,232.12) .. (260.38,230.67) -- cycle ;

\draw (43,184.4) node [anchor=north west][inner sep=0.75pt]    {$\B^n$};
\draw (230,66.4) node [anchor=north west][inner sep=0.75pt]  {$\text{bihol}^m\; \Phi$};
\draw (140,110) node [anchor=north west][inner sep=0.75pt]   {A linear fractional transformation:};
\draw (-5,211.4) node [anchor=north west][inner sep=0.75pt] {(strongly convex)};
\draw (500,211.4) node [anchor=north west][inner sep=0.75pt] {(weakly convex)};
\draw (441,85.4) node [anchor=north west][inner sep=0.75pt]    {$\Omega_0$};
\end{tikzpicture}}
\ee

In the figure above, we map a ball by a biholomorphic linear fractional transformation\footnote{Geometrically, the map $\Phi$ can be seen as a change of affine chart for $\mathbb{CP}^n$. This is what is indicated in the figure under the arrow in \eqref{cayley-figure} above. We think of the copy of $\mathbb{C}^n$ containing the ball on the left hand side of the biholomorphism above as the standard affine chart for complex projective space (the horizontal ``plane'' drawn in red in the figure below the arrow) so that $\mathbb{B}^n$ is identified with a subset of $\mathbb{CP}^n$ (a quadric, corresponding to the lines inside the cone in the picture below the arrow). The map $\Phi$ then simply amounts to viewing the ball $\mathbb{B}^n\subseteq \mathbb{CP}^n$ in a different affine chart for $\mathbb{CP}^n$ (corresponding to the ``plane'' drawn in blue that makes a $45^{\circ}$ angle to the first, which we think of as the copy of $\mathbb{C}^n$ on the right hand side of the biholomorphism arrow that contains $\Omega_0$). Note that $\Phi$ sends one point on $S^{2n-1}=\partial \mathbb{B}^n$ to infinity, namely the point
$(0,\ldots, 0,-1)$; this corresponds to the fact that the blue affine chart (the one drawn at $45^{\circ}$) cuts every line in the cone except the one corresponding to that point. The cone slicing picture very succinctly describes the generalized Cayley transform and shows the ``parabolic'' nature of the domain $\Omega_0$ (slicing a cone at $45^{\circ}$ gives a parabola). Unfortunately, however, we cannot draw in high enough dimensions to show how the flat direction in $\partial \Omega_0$ arises from slicing the ``cone'' in $\mathbb{C}^{n+1}$.} to the ``Siegel upper half-space'' $\Omega_0$; this map (or its inverse) is known as the \textit{(generalized) Cayley transform} as it generalizes the well-known Cayley map from the disc to the upper half-plane in one complex variable. Explicitly, the biholomorphism $\Phi:\mathbb{C}^n\setminus\{z_n=-1\} \to \mathbb{C}^n$ is defined by
$$
\Phi(z) = i\,\frac{e_n-z}{1+z_n}
$$
where $e_n=(0,\ldots, 0,1)$.
(In the $n=1$ case the map becomes $z\mapsto \displaystyle i\frac{1-z}{1+z}$, which is the inverse of the map $z\to \displaystyle\frac{i-z}{i+z}$ that takes the upper half plane to the disc.) Note that the domain on the left-hand side (the ball) is strongly convex, whereas the domain on the right-hand side ($\Omega_0$) is only convex.
If we perturb $\B^n$ slightly, it will remain strongly convex, but if we perturb $\Omega_0$ slightly, it may become non-convex.

If we bend the ``Siegel Upper Half Space" $\Omega_0=\{\mathrm{Im}\,w>\norm{z}^2\}$ as described above slightly to a domain $\Omega$ described by $\mathrm{Im}\,w>\varphi(z, \bar{z},\mathrm{Re}\,w)$ where the function $\varphi$ is defined by $\varphi(z, \bar{z},\mathrm{Re}\,w)=\norm{z}^2-\epsilon(\mathrm{Re}\,w)^2$ for points near the origin (with $\epsilon>0$ small) and is smoothly extended so that $\varphi(z, \bar{z},\mathrm{Re}\,w)=\norm{z}^2$ for points far from the origin, it remains a domain of holomorphy (assuming that the first and second derivatives of $\varphi$ behave mildly on the interpolating region, which can easily be ensured). It stays convex in the directions $z_1, z_2, \ldots, z_n$, but not in the $\mathrm{Re}\,w$ direction: 

\be
\adjustbox{valign=c,scale={0.9}{0.9}}{\tikzset{every picture/.style={line width=0.75pt}} 
\begin{tikzpicture}[x=0.75pt,y=0.75pt,yscale=-1,xscale=1]
\draw  [draw opacity=0] (471.01,129.8) .. controls (467,158.9) and (449.53,188.5) .. (431.96,195.95) .. controls (431.39,196.19) and (430.82,196.41) .. (430.25,196.6) -- (439.18,143.19) -- cycle ; \draw   (471.01,129.8) .. controls (467,158.9) and (449.53,188.5) .. (431.96,195.95) .. controls (431.39,196.19) and (430.82,196.41) .. (430.25,196.6) ;  
\draw  [draw opacity=0] (430.25,196.6) .. controls (427.45,197.41) and (424.59,197.54) .. (421.72,196.9) .. controls (409.33,194.13) and (400.66,177.72) .. (399.17,157) -- (428.94,144.15) -- cycle ; \draw   (430.25,196.6) .. controls (427.45,197.41) and (424.59,197.54) .. (421.72,196.9) .. controls (409.33,194.13) and (400.66,177.72) .. (399.17,157) ;  
\draw  [draw opacity=0] (272.25,195.6) .. controls (269.45,196.41) and (266.59,196.54) .. (263.72,195.9) .. controls (251.33,193.13) and (242.66,176.72) .. (241.17,156) -- (270.94,143.15) -- cycle ; \draw   (272.25,195.6) .. controls (269.45,196.41) and (266.59,196.54) .. (263.72,195.9) .. controls (251.33,193.13) and (242.66,176.72) .. (241.17,156) ;  
\draw  [draw opacity=0] (313.01,128.8) .. controls (312.31,133.89) and (311.2,138.99) .. (309.75,143.99) -- (281.18,142.19) -- cycle ; \draw   (313.01,128.8) .. controls (312.31,133.89) and (311.2,138.99) .. (309.75,143.99) ;  
\draw  [draw opacity=0][dash pattern={on 4.5pt off 4.5pt}] (308.23,148.76) .. controls (300.93,170.46) and (287.44,189.36) .. (273.9,195.1) .. controls (273.32,195.35) and (272.75,195.57) .. (272.19,195.76) -- (281.11,142.35) -- cycle ; \draw  [dash pattern={on 4.5pt off 4.5pt}] (308.23,148.76) .. controls (300.93,170.46) and (287.44,189.36) .. (273.9,195.1) .. controls (273.32,195.35) and (272.75,195.57) .. (272.19,195.76) ;  
\draw[<-]    (350,88.67) -- (350,146) node[pos=0,right]{$\text{Im}(\omega)$};
\draw  [dash pattern={on 0.84pt off 2.51pt}]   (350,146) -- (350,180) ;
\draw  [dash pattern={on 0.84pt off 2.51pt}]  (350,180) -- (405,180.5) ;
\draw[->]    (405,180.5) -- (472,180.5) node[pos=1,below]{$\text{Re}(\omega)$};
\draw  [dash pattern={on 0.84pt off 2.51pt}]  (350,180) -- (344,187) ;
\draw[<-]    (290,228.5) -- (345,185) node[right,pos=0,xshift=3]{$z_{1}\ldots z_{n-1} \qquad (\text{Im}(\omega)>\|z\|^2-\varepsilon(\text{Re}(\omega))^2)$};
\draw  [draw opacity=0] (270.54,195.99) .. controls (281.41,189.6) and (310.65,185.02) .. (345,185.02) .. controls (380,185.02) and (409.69,189.77) .. (420.06,196.35) -- (345,201.51) -- cycle ; \draw   (270.54,195.99) .. controls (281.41,189.6) and (310.65,185.02) .. (345,185.02) .. controls (380,185.02) and (409.69,189.77) .. (420.06,196.35) ;  
\draw  [draw opacity=0] (241.12,155.67) .. controls (252.93,149.41) and (283.37,144.95) .. (319.05,144.95) .. controls (357.21,144.95) and (389.37,150.05) .. (399.17,157) -- (319.05,161.45) -- cycle ; \draw   (241.12,155.67) .. controls (252.93,149.41) and (283.37,144.95) .. (319.05,144.95) .. controls (357.21,144.95) and (389.37,150.05) .. (399.17,157) ;  
\draw  [draw opacity=0] (313.01,128.8) .. controls (324.82,122.54) and (355.26,118.09) .. (390.95,118.09) .. controls (429.11,118.09) and (461.27,123.18) .. (471.07,130.13) -- (390.95,134.58) -- cycle ; \draw   (313.01,128.8) .. controls (324.82,122.54) and (355.26,118.09) .. (390.95,118.09) .. controls (429.11,118.09) and (461.27,123.18) .. (471.07,130.13) ;  
\draw (266,126.4) node [anchor=north west][inner sep=0.75pt]    {$\Omega$};
\end{tikzpicture}}
\ee

This is a non-convex example of a domain of holomorphy.
Although it is not convex, one can still find complex hypervariety tangent to each point of $\partial \Omega$ that otherwise lies on the outside. For example, we can use the Cayley transformation to view this as a perturbation of $\B^n$ (which is convex) and then carry the singular functions over to handle the domain $\Omega$.

The above discussion means that what we really want is some kind of a notion of convexity ``in the complex tangent directions.'' But only roughly: this would lead us to the notion of $\C$-linear convexity which is stronger than what we really want. We also want a notion of ``convexity'' that is invariant under (local) biholomorphic changes of coordinates. Before we come to the appropriate definition we will first consider some non-domains of holomorphy. A domain fails to be convex if there are two points in the domain such that the line segment between these two points contains points outside of the domain. We will see that a domain fails to be a domain of holomorphy if there is a complex disc (or more generally the image of a closed analytic disc $\varphi :\overline{\D}\to\C^n$) whose boundary lies inside the domain but whose interior contains points outside the domain, provided this disc can be obtained as a member of a continuous family of discs where one of the (closed) discs in the family is completely contained in the domain.

\subsection{Non-domains of holomorphy}

Now that we have encountered a number of domains of holomorphy, it is time to ask when does a domain fail to be one?
As a concrete example, consider the following \emph{Hartogs figure} $H \subseteq \C^2$ (with $0 <a,b <1$):
\be
\newcommand{\hartogstext}{\parbox[t]{2.3in}{In diagrams, the Hartogs figure is
often drawn as:}}
\begin{adjustbox}{valign=c}
\scalebox{1}{
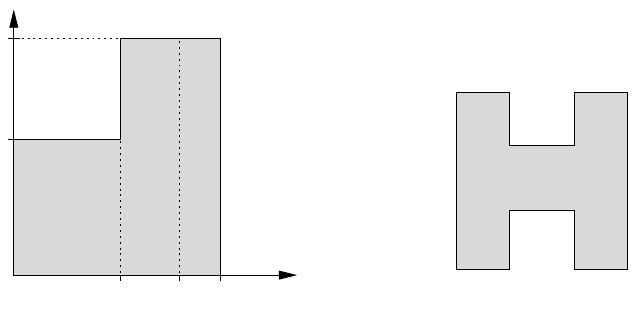
}
\end{adjustbox}
\ee
In drawing the diagram on the right, we have to imagine that $H$ revolves around in $(z,w)$ space with the radii given by the left picture. In particular, the ``arms'' of the ``H'' are connected to each other and contain circles that surround the white space between the arms. These circles bound a family of discs that cover all the points corresponding to the white space between the arms, and this is the key to the following theorem.
\begin{mtheorem}[Hartogs, 1906]
Every $f \in \sO(H)$ extends to $F \in \sO(\D \times \D)$.\\

\noindent\emph{Proof}:
For every $f \in \sO(H)$, we can use Cauchy's integral formula in $z$ to extend it to $F \in \sO(\D \times \D)$ using the analytic dependence of boundary values on $w$.
\end{mtheorem} 
Hence, $H$ is not a domain of holomorphy. 
For example, if we tried to put a pole somewhere in the middle, it would always poke out through the $H$. Hence, a pole contained in $\D \times \D\setminus H$ cannot exist. If we hypothetically had an isolated singularity, we could avoid $H$. Hence, a holomorphic function $F$ cannot have isolated singularities.

There are higher-dimensional versions of the above figure with $w = (w_1, \ldots, w_k)$ and $z = (z_1, \ldots, z_n)$. Moreover, we can dilate and rotate this example using unitary transformations to ``fit'' it near the boundaries of other domains. For example, one can use it to show that $\C^2 \setminus \R^2$ is not a domain of holomorphy, but $\C^2 \setminus (\C \times \{0\})$ is. In another special case, we have the following result on removable singularities.
\begin{mdcorollary}
For $n>1$, every $f \in \sO(\C^n \setminus \{0\})$ extends to $F \in \sO(\C^n)$. 
\end{mdcorollary}

\begin{QA}
If we have $U$ that is a domain of holomorphy but not convex, can we make a biholomorphic mapping to find a convex domain?\\

\noindent\emph{Answer:} Not in general. It is a very hard problem. We'll see that domains of holomorphy are characterized by pseudoconvexity  (meaning that the Levi form, which will be defined below, is nonnegative at every point). It is easy to show that a domain that is \textit{strongly pseudoconvex} (meaning that the Levi form is everywhere positive definite) can be locally made strongly convex by a biholomorphic change of coordinates, but this can not always be done globally. On the other hand, even locally not every pseudoconvex domain can be made convex by a biholomorphic change of coordinates, see \cite{KohnNirenberg1973,Kolar2010}.
\end{QA}

The key idea behind the Hartogs extension phenomenon described above is that one has a family of discs whose boundaries stay in the domain $U$ but whose interiors sweep out the region to which one is trying to extend. The next theorem, the Kontinuit\"atssatz, is essentially saying that to establish the local holomorphic extension of functions $f\in \mathcal{O}(U)$ accross a boundary point $p\in \partial U$ it is enough to find a family of holomorphic discs that remain inside $U$ (and whose boundaries stay ``well inside'') tending to a limiting disc that contains $p$ in its interior.
\begin{mtheorem}
    (Kontinuit\"atssatz, version 1)
    Suppose that $U \subseteq \C^n$ is open and there exists a sequence of closed analytic discs $\varphi_k: \overline{\D} \to \C^n$ converging (pointwise) to a closed analytic disc $\varphi: \overline{\D} \to \C^n$, such that $\varphi_k(\overline{\D}) \subseteq U$ for all $k$ and $\varphi(\partial\D) \subseteq U$. Then there exists a polyradius $s > 0$ such that for every $f \in \sO(U)$ and every $p \in \varphi(\D)$, there is an $F \in \sO(\Delta_s(p))$ with $F=f$ on some open subset of $U \cap \Delta_s(p)$.
\be
\adjustbox{valign=c,scale={1.3}{1.3}}{\tikzset{every picture/.style={line width=0.75pt}} 
\begin{tikzpicture}[x=0.75pt,y=0.75pt,yscale=-1,xscale=1]
\draw  [color=Gray  ,draw opacity=1 ][fill=Gray  ,fill opacity=0.2 ] (245.2,72.02) .. controls (262.23,88.99) and (253.33,141.01) .. (233.29,121.04) .. controls (213.26,101.08) and (193.26,103.12) .. (173.29,121.16) .. controls (153.33,139.2) and (133.24,93.24) .. (163.2,73.18) .. controls (193.16,53.12) and (228.17,55.05) .. (245.2,72.02) -- cycle ;
\draw    (170,83) -- (231,76) ;
\draw [shift={(231,76)}, rotate = 353.45] [color=black  ][fill=black  ][line width=0.75]      (0, 0) circle [x radius= 1.34, y radius= 1.34]   ;
\draw [shift={(170,83)}, rotate = 353.45] [color=black  ][fill=black  ][line width=0.75]      (0, 0) circle [x radius= 1.34, y radius= 1.34]   ;
\draw    (171,89) -- (232,82) ;
\draw [shift={(232,82)}, rotate = 353.45] [color=black  ][fill=black  ][line width=0.75]      (0, 0) circle [x radius= 1.34, y radius= 1.34]   ;
\draw [shift={(171,89)}, rotate = 353.45] [color=black  ][fill=black  ][line width=0.75]      (0, 0) circle [x radius= 1.34, y radius= 1.34]   ;
\draw    (172,94) -- (233,87) ;
\draw [shift={(233,87)}, rotate = 353.45] [color=black  ][fill=black  ][line width=0.75]      (0, 0) circle [x radius= 1.34, y radius= 1.34]   ;
\draw [shift={(172,94)}, rotate = 353.45] [color=black  ][fill=black  ][line width=0.75]      (0, 0) circle [x radius= 1.34, y radius= 1.34]   ;
\draw    (169,77) -- (230,70) ;
\draw [shift={(230,70)}, rotate = 353.45] [color=black  ][fill=black  ][line width=0.75]      (0, 0) circle [x radius= 1.34, y radius= 1.34]   ;
\draw [shift={(169,77)}, rotate = 353.45] [color=black  ][fill=black  ][line width=0.75]      (0, 0) circle [x radius= 1.34, y radius= 1.34]   ;
\draw    (172,97) -- (233,90) ;
\draw [shift={(233,90)}, rotate = 353.45] [color=black  ][fill=black  ][line width=0.75]      (0, 0) circle [x radius= 1.34, y radius= 1.34]   ;
\draw [shift={(172,97)}, rotate = 353.45] [color=black  ][fill=black  ][line width=0.75]      (0, 0) circle [x radius= 1.34, y radius= 1.34]   ;
\draw    (172,100) -- (233,93) ;
\draw [shift={(233,93)}, rotate = 353.45] [color=black  ][fill=black  ][line width=0.75]      (0, 0) circle [x radius= 1.34, y radius= 1.34]   ;
\draw [shift={(172,100)}, rotate = 353.45] [color=black  ][fill=black  ][line width=0.75]      (0, 0) circle [x radius= 1.34, y radius= 1.34]   ;
\draw    (172,104) -- (233,97) ;
\draw [shift={(233,97)}, rotate = 353.45] [color=black  ][fill=black  ][line width=0.75]      (0, 0) circle [x radius= 1.34, y radius= 1.34]   ;
\draw [shift={(172,104)}, rotate = 353.45] [color=black  ][fill=black  ][line width=0.75]      (0, 0) circle [x radius= 1.34, y radius= 1.34]   ;
\draw    (172,102) -- (233,95) ;
\draw [shift={(233,95)}, rotate = 353.45] [color=black  ][fill=black  ][line width=0.75]      (0, 0) circle [x radius= 1.34, y radius= 1.34]   ;
\draw [shift={(172,102)}, rotate = 353.45] [color=black  ][fill=black  ][line width=0.75]      (0, 0) circle [x radius= 1.34, y radius= 1.34]   ;
\draw    (172,107) -- (233,100) ;
\draw [shift={(233,100)}, rotate = 353.45] [color=black  ][fill=black  ][line width=0.75]      (0, 0) circle [x radius= 1.34, y radius= 1.34]   ;
\draw [shift={(172,107)}, rotate = 353.45] [color=black  ][fill=black  ][line width=0.75]      (0, 0) circle [x radius= 1.34, y radius= 1.34]   ;
\draw    (172,109) -- (233,102) ;
\draw [shift={(233,102)}, rotate = 353.45] [color=black  ][fill=black  ][line width=0.75]      (0, 0) circle [x radius= 1.34, y radius= 1.34]   ;
\draw [shift={(172,109)}, rotate = 353.45] [color=black  ][fill=black  ][line width=0.75]      (0, 0) circle [x radius= 1.34, y radius= 1.34]   ;
\draw [color=Maroon  ,draw opacity=1 ]   (172,110) -- (233,103) ;
\draw [shift={(233,103)}, rotate = 353.45] [color=Maroon  ,draw opacity=1 ][fill=Maroon  ,fill opacity=1 ][line width=0.75]      (0, 0) circle [x radius= 1.34, y radius= 1.34]   ;
\draw [shift={(172,110)}, rotate = 353.45] [color=Maroon  ,draw opacity=1 ][fill=Maroon  ,fill opacity=1 ][line width=0.75]      (0, 0) circle [x radius= 1.34, y radius= 1.34]   ;
\draw  [draw opacity=0][fill=RoyalBlue  ,fill opacity=1 ] (204.5,107.5) .. controls (204.5,106.4) and (203.6,105.5) .. (202.5,105.5) .. controls (201.4,105.5) and (200.5,106.4) .. (200.5,107.5) .. controls (200.5,108.6) and (201.4,109.5) .. (202.5,109.5) .. controls (203.6,109.5) and (204.5,108.6) .. (204.5,107.5) -- cycle ;
\draw (154,91.4) node [anchor=north west][inner sep=0.75pt, scale=0.8]  [color=Gray  ,opacity=1 ]  {$U$};
\draw (237,60.4) node [anchor=north west][inner sep=0.75pt, scale=0.8]    {$\varphi _{k}(\D)$};
\draw (237,100.4) node [anchor=north west][inner sep=0.75pt]  [color=Maroon  ,opacity=1, scale=0.8]  {$\varphi (\D)$};
\draw (204.5,111.9) node [anchor=north west][inner sep=0.75pt,scale=0.8]  [color=RoyalBlue  ,opacity=1 ]  {$p$};
\end{tikzpicture}}
\ee
In the above figure the theorem gives us a fixed neighborhood $\Delta_s(p)$ of the point $p$ such that every holomorphic function $f$ in $U$ extends to $\Delta_s(p)$.
\end{mtheorem}
The idea behind the proof is as follows. The boundary points of the discs stay bounded away from $\partial U$ and the Cauchy estimates at/near the boundary points propagate to the interior of the discs by the maximum principle. Thus, uniform lower bounds for the ``polyradii'' of convergence are obtained, giving the result.

Now, we want to find out what is the right condition on $\partial U$ (assuming that it is smooth) to avoid the presence of the above discs. This question can be studied with the Cauchy--Riemann (CR) structure of the boundary.

\subsection{The CR geometry of the boundary}

Consider a local biholomorphism $\Phi$ from $M$ to some $M'$. Which part of the geometry is preserved? 

\be
\adjustbox{valign=c,scale={0.9}{0.9}}{\tikzset{every picture/.style={line width=0.75pt}} 
\begin{tikzpicture}[x=0.75pt,y=0.75pt,yscale=-1,xscale=1]
\draw    (79.34,110) .. controls (114,118) and (134,169) .. (128.87,179) ;
\draw    (128.87,179) .. controls (165,162) and (217,167) .. (234.66,179) ;
\draw    (185.13,110) .. controls (209,126) and (225,152) .. (234.66,179) ;
\draw    (79.34,110) .. controls (116,102) and (146,98) .. (185.13,110) ;
\draw    (355,124) .. controls (395,94) and (416,123) .. (456,93) ;
\draw    (374,175) .. controls (414,145) and (435,174) .. (475,144) ;
\draw    (355,124) -- (374,175) ;
\draw    (456,93) -- (475,144) ;
\draw  [draw opacity=0] (197.58,104.22) .. controls (214.67,95.63) and (245.08,89.78) .. (279.8,89.54) .. controls (315.8,89.3) and (347.26,95.15) .. (364.06,104.04) -- (280,119.54) -- cycle ; \draw[->]   (197.58,104.22) .. controls (214.67,95.63) and (245.08,89.78) .. (279.8,89.54) .. controls (315.8,89.3) and (347.26,95.15) .. (364.06,104.04) ;  
\draw  [draw opacity=0][fill={rgb, 255:red, 0; green, 0; blue, 0 }  ,fill opacity=1 ] (159.38,136.67) .. controls (159.38,135.22) and (160.55,134.04) .. (162,134.04) .. controls (163.45,134.04) and (164.63,135.22) .. (164.63,136.67) .. controls (164.63,138.12) and (163.45,139.29) .. (162,139.29) .. controls (160.55,139.29) and (159.38,138.12) .. (159.38,136.67) -- cycle ;
\draw  [draw opacity=0][fill={rgb, 255:red, 0; green, 0; blue, 0 }  ,fill opacity=1 ] (413.38,133.67) .. controls (413.38,132.22) and (414.55,131.04) .. (416,131.04) .. controls (417.45,131.04) and (418.63,132.22) .. (418.63,133.67) .. controls (418.63,135.12) and (417.45,136.29) .. (416,136.29) .. controls (414.55,136.29) and (413.38,135.12) .. (413.38,133.67) -- cycle ;
\draw  [dash pattern={on 4.5pt off 4.5pt}] (374,133.67) .. controls (374,110.47) and (392.8,91.67) .. (416,91.67) .. controls (439.2,91.67) and (458,110.47) .. (458,133.67) .. controls (458,156.86) and (439.2,175.67) .. (416,175.67) .. controls (392.8,175.67) and (374,156.86) .. (374,133.67) -- cycle ;
\draw  [dash pattern={on 4.5pt off 4.5pt}] (120,139.29) .. controls (120,116.1) and (138.8,97.29) .. (162,97.29) .. controls (185.2,97.29) and (204,116.1) .. (204,139.29) .. controls (204,162.49) and (185.2,181.29) .. (162,181.29) .. controls (138.8,181.29) and (120,162.49) .. (120,139.29) -- cycle ;
\draw (154,187.4) node [anchor=north west][inner sep=0.75pt]    {$M$};
\draw (226,60.4) node [anchor=north west][inner sep=0.75pt]    {local $\text{bihol}^m~ \Phi$};
\end{tikzpicture}}
\ee

The derivative $D \Phi$ is a $\C$-linear isomorphism at each point. It preserves the \textit{maximal complex subspace} $H_p \subseteq T_p M$ and the \textit{complex structure} $J_p: H_p \to H_p$,  the restriction of the ambient complex structure $\mathbb{J}_p:T_p\C^n\to T_p\C^n$ to $H_p=T_pM\cap \mathbb{J}_pT_pM$. Since $\mathbb{J}_p^2 = - \mathrm{Id}$ on $T_p\C^n$,  $J_p^2 = - \mathrm{Id}$ on $H_p$. Vectors not lying in $H_p$ get rotated out of the tangent space $T_p M$ when acted on by $\mathbb{J}_p$: 
\be
\adjustbox{valign=c,scale={1.3}{1.3}}{\tikzset{every picture/.style={line width=0.75pt}}
\begin{tikzpicture}[x=0.75pt,y=0.75pt,yscale=-0.9,xscale=0.9]
\draw    (245.34,118) .. controls (280,126) and (300,177) .. (294.87,187) ;
\draw    (294.87,187) .. controls (331,170) and (383,175) .. (400.66,187) ;
\draw    (351.13,118) .. controls (375,134) and (391,160) .. (400.66,187) ;
\draw    (245.34,118) .. controls (282,110) and (312,106) .. (351.13,118) ;
\draw[color=Maroon]    (296.5,147.67) -- (359.5,141.67) ;
\draw  [draw opacity=0][fill={rgb, 255:red, 0; green, 0; blue, 0 }  ,fill opacity=1 ] (325.38,144.67) .. controls (325.38,143.22) and (326.55,142.04) .. (328,142.04) .. controls (329.45,142.04) and (330.63,143.22) .. (330.63,144.67) .. controls (330.63,146.12) and (329.45,147.29) .. (328,147.29) .. controls (326.55,147.29) and (325.38,146.12) .. (325.38,144.67) -- cycle ;
\draw  [draw opacity=0] (316.19,143.57) .. controls (315.44,142.49) and (314.87,141.24) .. (314.55,139.88) .. controls (313.14,133.87) and (317.09,127.81) .. (323.36,126.34) .. controls (329.64,124.87) and (335.87,128.55) .. (337.28,134.56) .. controls (337.59,135.92) and (337.64,137.28) .. (337.44,138.6) -- (325.91,137.22) -- cycle ; \draw[color=RoyalBlue,<-]   (316.19,143.57) .. controls (315.44,142.49) and (314.87,141.24) .. (314.55,139.88) .. controls (313.14,133.87) and (317.09,127.81) .. (323.36,126.34) .. controls (329.64,124.87) and (335.87,128.55) .. (337.28,134.56) .. controls (337.59,135.92) and (337.64,137.28) .. (337.44,138.6) ;  
\draw (295.5,151.07) node [anchor=north west,color=Maroon,scale=0.8][inner sep=0.75pt]    {$H_p$};
\draw (295.5,120.07) node [anchor=north west,color=RoyalBlue,scale=0.8][inner sep=0.75pt]    {$\mathbb{J}_p$};
\end{tikzpicture}}
\ee
So the part of the geometry that is preserved is the \emph{CR structure}: $(M,H,J)$.
Although $J_p$ allows us to think of the real $2(n-1)$ dimensional vector space $H_p$ as a complex vector space (defining the multiplication by $i$ by $iX =J_p X$ for $X\in H_p$) it is customary and in the end much more convenient to think of $H_p$ as being a real vector space. Since the operator $J_p: H_p \to H_p$ squares to minus the identity, it has eigenvalues $\pm i$; hence, if we complexify  $H_p$ to $\C H_p = \C\otimes_{\mathbb{R}}H_p$ we may decompose $\C H_p$ into $i$ and $-i$ eigenspaces as $\C H_p = T_p^{(1,0)}M \oplus T^{(0,1)}_p M$. Here, $T_p^{(1,0)}M$ is the $(n-1)$-dimensional complex vector space $\{ X-iJ_pX\,|\, X\in H_p\}$ and $T_p^{(0,1)}M$ is the conjugate vector space $\{ X+iJ_pX\,|\, X\in H_p\}$. Equivalently, $T^{1,0}_p M =  T^{(1,0)}_p \C^n \cap \C T_p M = \mathrm{ker}\, \partial r |_p$ where $r$ is a local defining function for $M$ and 
\be
T^{(1,0)}_p \C^n = \left\{\left.\frac{\partial}{\partial z_1}\right|_p, \ldots , \left.\frac{\partial}{\partial z_n}\right|_p \right \}.
\ee
Knowing $T_p^{(1,0)}M$ is equivalent to knowing $H_p$ and $J_p$ so this gives an alternative way to define the CR structure. For convenience we will write $d=n-1$ for the complex dimension of $T_p^{(1,0)}M$, which is known as the \textit{CR dimension} of $M$. The space $T_p^{(1,0)}M$ is known as the \textit{holomorphic tangent space} to $M$ at $p$. Typically $T_p^{(1,0)}M$ is described by giving an explicit frame, see, e.g., Examples \ref{ex:Heisenberg-in-C3} and \ref{ex:indefinite-Levi} in Lecture 4.

The key point is this. If we pick a complementary direction $T$ (a vector field tangent to $M$) to $H_p \subseteq T_p M$, then the direct sum decomposition
\be
T_p M = H_p \oplus \mathbb{R} T_p\,,
\ee
complexifies to
\be\label{eq:complex-tangent-decomp}
\C T_p M = T_p^{(1,0)}M \oplus T_p^{(0,1)}M \oplus \C T_p\, .
\ee
The final decomposition can be written as $\mathcal{E}^{\alpha} \oplus \mathcal{E}^{\overline{\alpha}} \oplus \mathcal{E}^0$ in Penrose-style notation.
If $r$ is a local defining function for $M=\partial U$ (meaning that $r < 0$ in $U$ while $r = 0$ on $\partial U$ and $d r \neq 0$ on $\partial U$), we can write the corresponding decomposition of the tangential part of its Hessian using Penrose abstract index notation as:
\be\label{eq:tangential-hessian-decomp}
\text{Hess}(r) \big|_{T_p M} = \begin{pmatrix}
r_{\alpha\beta} & r_{\alpha \bar{\beta}} & r_{\alpha 0}\\
r_{\bar{\alpha}\beta} & r_{\bar{\alpha}\bar{\beta}} & r_{\bar{\alpha}0}\\
r_{0\beta} & r_{0\bar{\beta}} & r_{00}
\end{pmatrix} .
\ee
Here, $r_{\alpha\overline{\beta}}$ and $r_{\overline{\alpha}\beta}$ are conjugates, as are $r_{\alpha\beta}$ and $r_{\overline{\alpha}\overline{\beta}}$. In this notation, convexity means that this Hessian is positive (in the tangential directions):
\be
\text{Hess}(r) \big|_{T_p M} \geq 0 \quad\text{for all }p\in M\,.
\ee
On the other hand, pseudoconvexity means that only
\be\label{ineq:pseudoconvex}
r_{\alpha \overline{\beta}} \geq 0 \quad\text{for all }p\in M\, .
\ee

\begin{mdremark}
The Penrose-style notation is meant to be interpreted as expressing the abstract block decomposition of $\text{Hess}(r) \big|_{T_p M}$ corresponding to the decomposition \eqref{eq:complex-tangent-decomp} of the (complexified) tangent space. But, of course, if one were to introduce a basis $(Z_1, \ldots,  Z_{d})$ for $T_p^{(1,0)} M$ and set $Z_{\bar{\alpha}} = \overline{Z_{\alpha}}$, then $(Z_1, \ldots , Z_{d}, Z_{\bar{1}}, \ldots , Z_{\bar{d}}, T)$ would be a basis of $\C T_p M$ that respects the decomposition \eqref{eq:complex-tangent-decomp} and \eqref{eq:tangential-hessian-decomp} could then be interpreted as giving the $(2d+1)\times (2d+1)$ matrix for $\text{Hess}(r) \big|_{T_p M}$; in this case we can interpret $r_{\alpha \overline{\beta}}$ as $\text{Hess}(r)(Z_{\alpha}, Z_{\bar{\beta}})$ and condition \eqref{ineq:pseudoconvex} is the condition that the matrix $(r_{\alpha \overline{\beta}})$ is positive semidefinite.
\end{mdremark}

Finally, let us wrap up this lecture with a precise definition.
\begin{mddefinition}[Levi pseudoconvexity]\label{def:levi-pseudoconvex}
Suppose $U \subseteq \C^n$ is an open set with smooth boundary, and $r$ is a defining function for $U$ at $p$ with $r<0$ in $U$. Then $U$ is said to be \textit{pseudoconvex} at $p\in M=\partial U$ if
    \be\label{ineq:pseudoconvex-def}
\sum_{j,k=1}^{n} a_j \bar{a}_k \frac{\partial^2 r}{\partial z_j \partial \bar{z}_k} \bigg|_p \geq 0 \qquad\text{for all }\quad X_p = \underbrace{\sum_{j=1}^{n} a_j \frac{\partial}{\partial z_j} \bigg|_{p} \in T_p^{(1,0)} M}_{X_p r = \sum_{j=1}^{n} a_j \frac{\partial r}{\partial z_j} \big|_p = 0}\, .
\ee
The hermitian form on $T_p^{(1,0)} M$ defined by the left hand side of the above display is called the \textit{Levi form} of $M=\partial U$ at $p$. So, a domain is said to be pseudoconvex at $p\in \partial U$ if and only if the Levi form of $\partial U$ is positive semidefinite at $p$.
\end{mddefinition}

This is the biholomorphically invariant part of the convexity condition. Given a point $p\in M$ (using $dr_p\neq 0$) one can find a local biholomorphic change of coordinates so that in the new coordinates $\zeta$ we have $\left.\frac{\partial^2 r}{\partial \zeta_j \partial \zeta_k}\right|_p = 0$. On the other hand, the \textit{complex Hessian} $\frac{\partial^2 r}{\partial z_j \partial \bar{z}_k}$ transforms as a tensor under change of coordinates: if $\zeta_m = f_m(z_1, \ldots, z_n)$, $m=1,\ldots , n$, is a local biholomorphic change of variables and $f=(f_1,\ldots, f_m)$ then from the chain rule and the fact that $\frac{\partial f_{m}}{\partial \bar{\zeta}_j}=0$ one can easily show that
\be
\frac{\partial^2 (r\circ f)}{\partial \zeta_j \partial \bar{\zeta}_k} =  \sum_{\ell, m = 1}^n \frac{\partial^2 r}{\partial z_{\ell} \partial \bar{z}_m} \frac{\partial f_{\ell}}{\partial \zeta_j}\frac{\partial \overline{f_{m}}}{\partial \bar{\zeta}_k}\,,
\ee
(the corresponding formula for $\frac{\partial^2 (r\circ f)}{\partial \zeta_j \partial \zeta_k}$ involves the second derivatives of $f$ and from the explicit expression one can see that at a point $p$ where $dr\neq 0$ one can find a change of coordinates map $f$ such that $\frac{\partial^2 (r\circ f)}{\partial \zeta_j \partial \zeta_k}$ is zero at $p$).

What is the significance in restricting the complex Hessian appearing in \eqref{ineq:pseudoconvex-def} to holomorphic tangent vectors $X_p\in T_p^{(1,0)}M$  to $M$ (rather than considering all $X_p\in T_p^{(1,0)}\C^n$)? The key point here is that $\sum_{j,k=1}^{n} a_j \bar{a}_k \left.\frac{\partial^2 r}{\partial z_j \partial \bar{z}_k} \right|_p$ only behaves nicely under a change in the defining function $r$ when  $X_p = \sum_{j=1}^{n} a_j \left.\frac{\partial}{\partial z_j} \right|_{p} \in T_p^{(1,0)} M$  (equivalently, when  $X_p r = \sum_{j=1}^{n} a_j \frac{\partial r}{\partial z_j} \big|_p =0$): If $\tilde{r}=\e^\Upsilon r$ is another defining function (where $\Upsilon$ is an arbitrary smooth real valued function) then, since $r(p)=0$,
\be
\sum_{j,k=1}^{n} a_j \bar{a}_k \frac{\partial^2 \tilde{r}}{\partial z_j \partial \bar{z}_k} \bigg|_p = \e^{\Upsilon}\sum_{j,k=1}^{n} \left( a_j \bar{a}_k \frac{\partial^2 r}{\partial z_j \partial \bar{z}_k}  + a_j \frac{\partial \Upsilon}{\partial z_j} a_k\frac{\partial r}{\partial \bar{z}_k} + a_j\frac{\partial r}{\partial z_j}a_k\frac{\partial \Upsilon}{\partial \bar{z}_k}\right)\bigg|_p.
\ee
When $X_p\in T_p^{(1,0)}M$ this simplifies to 
\be
\sum_{j,k=1}^{n} a_j \bar{a}_k \frac{\partial^2 \tilde{r}}{\partial z_j \partial \bar{z}_k} \bigg|_p = \e^{\Upsilon}\sum_{j,k=1}^{n} a_j \bar{a}_k \frac{\partial^2 r}{\partial z_j \partial \bar{z}_k}\bigg|_p\,,
\ee
and hence the condition \eqref{ineq:pseudoconvex-def} in Definition \ref{def:levi-pseudoconvex} does not depend on the choice of defininig function $r$. Moreover, the Levi form of $M=\partial U$ at $p$ is well defined (independent of the choice of the defining function) up to multiplication by a positive constant.

\begin{mdremark}
Some of the material covered at the start of Lecture 4 (Section \ref{sec:d-bar-problem}) was meant to be covered before Jiří's lecture on CR functions (Section \ref{sec:CR-functions}), but there was not enough time. The reader may find it helpful to look at these parts of Lecture 4 (Subsections \ref{subsec:Levi-revisited} and \ref{subsec:equivalences}) before reading Section \ref{sec:CR-functions}.
\end{mdremark}

\section[CR functions]{CR functions\label{sec:CR-functions}\\
\normalfont{\textit{Jiří Lebl}}}

\subsection{Real analytic functions and complexification}

Let us recall a simple result, which is a traditional way of interpreting complexification.

If $U \subset \C^n$ is a domain, $U \cap \R^n \not= \emptyset$,
$f,g \in \sO(U)$, and $f=g$ on $U \cap \R^n$, then $f \equiv g$.
The result goes the other way as well: If $V \subset \R^n$, $f \colon V \to \R$ is
real-analytic (that is, locally given by real power series), then
there exists an open domain $U \subset \C^n$ such that $V \subset U$ and a unique holomorphic function $F \in \sO(U)$ such that
$F|_V = f$.
The proof is essentially that given real power series
$\sum_{\alpha} c_n(x-p)^n$, we can
insert complex numbers and obtain
$\sum_{\alpha} c_n(z-p)^n$.

There is usually a better way to complexify real-analytic functions in $\C^n$.
Suppose $U \subset \C^n \cong \R^{2n}$ and $f \colon U \to \C$ is real-analytic. Write (at $0$ for simplicity)
\be
f(x,y)
=
\sum_{m=0}^\infty
f_m(x,y)
=
\sum_{m=0}^\infty
f_m\left(
\frac{z+\bar{z}}{2},
\frac{z-\bar{z}}{2i}\right)\,.
\ee
The polynomial $f_m$ becomes a homogeneous polynomial of degree $m$ in the variables $z$ and $\bar{z}$, which means the series is a power series in $z$ and $\bar{z}$. Hence, at any point, the function $f$ equals
\be
\sum_{\alpha,\beta} c_{\alpha,\beta} {(z-a)}^\alpha
{(\bar{z}-\bar{a})}^\beta\,,
\ee
in multinomial notation.
We will simply write it as $f(z,\bar{z})$. A holomorphic function is real-analytic, but not vice versa. A holomorphic function is a real-analytic function that does not depend on $\bar{z}$.

Before further discussion, we will need the following result. Let $U \subset \C^n \times \C^n$ be a domain and $f,g \in \sO(U)$
such that $f=g$ on the \emph{diagonal}
\be
U \cap D = U \cap \bigl\{ (z,\zeta) \in \C^n \times \C^n : \zeta = \bar{z} \bigr\}.
\ee
Then $f\equiv g$ on all of $U$.
The result also goes the other way: If $f \colon V \subset D \to \C$ is real-analytic,
then $f$ extends to a neighborhood of $V$ in $\C^{2n}$.
We identify $\C^n$ and $D \subset \C^n \times \C^n$ with $\iota(z) =
(z,\bar{z})$.

\begin{mdexample}
In one dimension, the function
\be
f(z,\bar{z}) = \frac{1}{1+\abs{z}^2} = \frac{1}{1+z\bar{z}}\,,
\ee
is
real-analytic in $\C$, but is not a restriction on the diagonal of a holomorphic function in all of $\C^2$. The problem is that the complexified function 
\be
f(z,\zeta) = \frac{1}{1+z\zeta}\,,
\ee
is
holomorphic in $\C^2 \setminus \{ z\zeta = -1 \}$, i.e., it is undefined on the set where $z\zeta = -1$.
\end{mdexample}

\begin{mdexample}
If $u(z,\bar{z})$ is a (pluri)harmonic function, then
$u(z,\bar{z}) = \Re f(z)$. How can we recover $f$ from $u$ in this case? First, notice that we have
\be
u(z,\bar{z}) = \dfrac{f(z)+\bar{f}(\bar{z})}{2}\, .
\ee
Without loss of generality, we can set $f(0)=0$. This implies
\be\label{eq:f-u}
f(z) = 2u(z,0)\, .
\ee
\end{mdexample}
The idea of treating $\bar{z}$ as a separate variable is very powerful, and as we have just seen, it is completely natural when speaking about real-analytic functions.\footnote{Jiří recalls a story of asking his advisor if there is a typo in the formula \eqref{eq:f-u} because it looks too good to be true.} This is one of the reasons why real-analytic functions play a special role in several complex variables.
\begin{mdremark}
There is no good control over the neighborhood to which $f$ extends.  This is true even in
1D: Given any interval $(a,b)$ and any neighborhood $U$ of
$(a,b)$, there is a function $F \in \sO(U)$ that does not extend beyond any boundary
point of $U$.  So $f = F|_{(a,b)}$ also cannot extend further. If we want any additional estimates to hold, we need to know something more about the function.
\end{mdremark}

\subsection{CR functions}

So far we have talked about the submanifold $\R^n \subset \C^n$. What can we say about more complicated submanifolds?

Suppose $M \subset \C^n$ is a hypersurface, then a function $f \colon M \to \C$
is a \emph{CR function} if
\be
X_p f = 0\,,
\ee
for all vectors $X_p \in T_p^{(0,1)} M$ for
all points $p \in M$.
Moreover, if $M \subset U \subset \C^n$ and $F \in \sO(U)$, then $F|_M$ is a CR function.

The question is whether the reverse statement holds. In fact, it is not always true: If $M$ is
real-analytic, then also $F|_M$ is real-analytic, so no smooth-only CR function $f$ on $M$ is such a restriction.
\begin{mtheorem}[Severi]
If $M$ and $f$ are real-analytic and $f$
CR, then $f$ extends holomorphically to a neighborhood.\\

\noindent\emph{Proof:} The proof feels like cheating, so let us do it.
Suppose without loss of generality that $0 \in M$ and $M$ is
real-analytic. Then, 
there is a holomorphic function
$\Phi(z,\zeta,w)$ in a neighborhood of $0$ in
$\C^{n-1} \times \C^{n-1} \times \C$,
such that $M$ is
\be
\bar{w} = \Phi(z,\bar{z},w) ,
\ee
and $\Phi$,
$\frac{\partial \Phi}{\partial z_k}$,
$\frac{\partial \Phi}{\partial \zeta_k}$
vanish at $0$
and $w = \bar{\Phi}\bigl(\zeta,z,\Phi(z,\zeta,w)\bigr)$.
A basis for $T^{(0,1)} M$ is given by
\be
\frac{\partial}{\partial \bar{z}_k}
+\frac{\partial \Phi}{\partial \bar{z}_k} \frac{\partial}{\partial \bar{w}}
\quad
\left(
=
\frac{\partial}{\partial \bar{z}_k}
+\frac{\partial \Phi}{\partial \zeta_k} \frac{\partial}{\partial \bar{w}}
\right)
,
\qquad k=1,\ldots,n-1.
\ee
We therefore conclude that $M$ is
$\bar{w} = \Phi(z,\bar{z},w)$,
$T^{(0,1)} M$ is given by
$\frac{\partial}{\partial \bar{z}_k}
+\frac{\partial \Phi}{\partial \bar{z}_k} \frac{\partial}{\partial
\bar{w}}$.
Define the complexification $\sM \subset \C^{2n}$ by
$\omega = \Phi(z,\zeta,w)$.
Moreover, let us complexify $f(z,w,\bar{z},\bar{w})$ to $f(z,w,\zeta,\omega)$.

Now comes the trick: Let us define
\be
F(z,w,\zeta) =
f\bigl(z,w,\zeta,\Phi(z,\zeta,w)\bigr) .
\ee
Because $f$ is a CR function, it is killed by
$\frac{\partial}{\partial \bar{z}_k}
+\frac{\partial \Phi}{\partial \bar{z}_k} \frac{\partial}{\partial
\bar{w}}$ on $M$. Hence
\be
\frac{\partial F}{\partial \zeta_k}
+\frac{\partial \Phi}{\partial \zeta_k} \frac{\partial F}{\partial
\omega}
=
\frac{\partial F}{\partial \zeta_k} = 0 .
\ee
This is true everywhere by complexification.
So $F$ is a function of $z$ and $w$ only. Therefore, $F$ is a holomorphic function on
$\C^n$ (on some neighborhood of $M$).  \hfill$\square$
\end{mtheorem}

\begin{mdexample}
Consider $M \subset \C^2$ given by
\be
\Im w = \sabs{z}^2\,.
\ee
That is,
$\frac{w-\bar{w}}{2i} = z \bar{z}$, or in other words, $\sM$
is given by
$\omega = -2i z\zeta + w$,
and the CR vector field by
$\frac{\partial}{\partial \bar{z}}
-2i z \frac{\partial}{\partial \bar{w}}$.
\end{mdexample}

The most important place where we find CR functions that are not necessarily real-analytic is as boundary values of holomorphic functions.
\begin{mdproposition}
Suppose $U \subset \C^n$ is open with smooth boundary and
$f \colon \widebar{U} \to \C$ is smooth and holomorphic on $U$.
Then $f|_{\partial U}$ is a smooth CR function.\\

\noindent\emph{Proof}: Each $X_p \in T_p^{(0,1)} \partial U$ is a limit of 
$T^{(0,1)} \C^n$ vectors from the inside. \hfill$\square$
\end{mdproposition}

The boundary values of a holomorphic function define the function uniquely. That is, if two holomorphic functions continuous up to the (smooth) boundary are equal on an open set of the boundary, then they are equal in the domain. This statement is made precise in the following proposition.
\begin{mdproposition}
Suppose $U \subset \C^n$ is a domain with smooth boundary and
$f \colon \widebar{U} \to \C$ is smooth, holomorphic on $U$
and $f|_{\partial U}$ is zero on a nonempty open subset.
Then $f \equiv 0$.\\

\noindent\emph{Proof:} Take $p \in \partial U$ such that $f=0$ on a neighborhood of $p$ in $\partial U$. Consider a small neighborhood $\Delta$ of $p$ such that $f$ is zero on $\partial U \cap \Delta$. Define $g: \Delta \to \C$ by setting $g(z) = f(z)$ if $z \in U$ and $g(z) = 0$ otherwise.
\be
\begin{adjustbox}{valign=c}
\scalebox{1}{
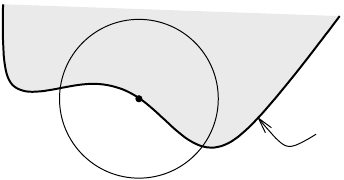
}
\end{adjustbox}
\ee
It is not hard to see that $g$ is continuous, and it is clearly holomorphic where is is not zero.
Use Rad{\'o}'s theorem (see below) to extend $g$
as $0$ outside,
then use the identity theorem. \hfill$\square$
\end{mdproposition}

\begin{mtheorem}[Rad\'o]
If $U \subset \C^n$ is open and
$g \colon U \to \C$ continuous
and holomorphic on
\begin{equation}
U' = \bigl\{ z \in U : g(z) \not= 0 \bigr\} .
\end{equation}
Then $g \in \sO(U)$.
\end{mtheorem}
This is Jiří's favorite theorem.

\subsection{Approximation of CR functions}

A problem we tackle next is trying to extend a smooth CR function from the boundary of a domain to a holomorphic function inside of it. This is essentially a PDE problem where the PDE are the Cauchy--Riemann equations, and the function on the boundary sets the boundary conditions. But notice that Cauchy--Riemann equations are \emph{overdetermined}, that is, there are too many equations. It means that not every boundary data gives a solution. The two propositions above say that, respectively, the boundary data being CR is a necessary condition for a solution (it is not sufficient in general) and that a solution is unique if it exists.

Let us give examples of functions that are not boundary values of holomorphic functions.
\begin{mdexample}
Suppose $M = \R \subset \C$.
Let us define the following function $f \colon M \to \C$:
\be
f(x) =
\begin{cases} 
\e^{-x^{-2}} & \text{if $x \not= 0$,} \\
0 & \text{if $x = 0$.}
\end{cases} 
\ee
Then $f$ is CR (trivially), but is not a restriction nor a boundary value (from
either side) of a
holomorphic function continuous up to $0$. We can come up with generalizations of this example to several variables by working on $M = \R \times \C$.
\end{mdexample}

\begin{mdexample}
Define the function $f \in \overline{\bB}_2 \to \C$ by
\be
f(z_1,z_2) =
\begin{cases} 
\e^{-1/\sqrt{z_1+1}} & \text{if $z_1 \not= -1$,} \\
0 & \text{if $z_1 = -1$.}
\end{cases} 
\ee
Then $f$ is smooth on $\overline{\bB}_2$, holomorphic on $\bB_2$,
but near $(-1,0)$ it is not a restriction of a holomorphic function
(it is only a one sided extension).
\end{mdexample}

A neat technique for extension is to approximate functions by polynomials. The following theorem holds in much more generality, but here we state its simplest version.
\begin{mtheorem}[Baouendi--Tr{\`e}ves]
Suppose $M \subset \C^n$ is a smooth real hypersurface,
$p \in M$.
Then there exists a compact
neighborhood $K \subset M$ of $p$, such that for every CR function
$f \colon M \to \C$,
there exists a sequence $\{ p_\ell \}$ of polynomials in $z$ such that
\begin{equation}
p_\ell(z) \to f(z)
\qquad \text{uniformly in $K$.}
\end{equation}
\end{mtheorem}
A key point is that $K$ cannot be chosen arbitrarily, as it depends on $p$ and $M$. On the other hand, it does not depend on $f$. Given $M$ and $p \in M$, there is a $K$ such that \emph{every} CR function on $M$ is approximated uniformly on $K$ by holomorphic polynomials.
\begin{mdexample}
The $K$ depends only on $M$, but can not always be all of $M$:
For example, $M = S^1$ and $f = \bar{z}$.
\end{mdexample}

The proof is based on the standard proof of Weierstra\ss\, theorem:
If $f \colon [0,1] \to
\R$ is continuous, then it is approximated on $[0,1]$ by the entire functions
\be
f_\ell(z) = \int_0^1 c_\ell\, \e^{-\ell (z-t)^2} f(t) \, \d t\,,
\ee
for properly chosen $c_\ell$. The proof involves taking partial sums of the power series.
Baouendi--Tr{\`e}ves uses the same idea on a totally real subset of $M$ and a slightly modified version of the above argument.

\subsection{Extension of CR functions}

The following is called
the Lewy extension theorem, but goes back to
Helmut Knesser in 1936.
\begin{mtheorem}
Suppose $M \subset \C^n$ is a smooth real hypersurface and $p \in M$.
There exists a neighborhood $U$ of $p$ with the following
property.
Suppose $r \colon U \to \R$ is
a smooth defining function for $M \cap U$, denote by $U_- \subset U$ the set where $r$
is negative and $U_+ \subset U$ the set where $r$ is positive.
Let $f \colon M \to \R$ be a smooth CR function.
Then:
\begin{enumerate}[label=(\roman*)]
\item
If the Levi form with respect to $r$ has a positive eigenvalue at $p$, then
$f$ extends to a holomorphic function on $U_-$ continuous up to $M$.
\item
If the Levi form with respect to $r$ has a negative eigenvalue at $p$, then
$f$ extends to a holomorphic function on $U_+$ continuous up to $M$.
\item
If the Levi form with respect to $r$ has eigenvalues of both signs at $p$, then
$f$ extends to a holomorphic function on $U$.
\end{enumerate}
\end{mtheorem}
So, if the Levi-form has eigenvalues of both signs,
then every CR function is a restriction of a holomorphic function.

Let us provide a quick sketch of the proof of (i). First, suppose $p=0$ and write $M$ in the neighborhood of the origin as
\be
\Im w = \sabs{z_1}^2 + \sum_{k=2}^{n-1} \epsilon_k \sabs{z_k}^2 +
E(z_1,z',\bar{z}_1,\bar{z}',\Re w) ,
\ee
where $z' = (z_2,\ldots,z_{n-1})$, $\epsilon_k = -1,0,1$, and $E$ is $O(3)$.
Next, apply Bauoendi--Tr{\`e}ves theorem to find a compact neighborhood $K$ of the origin of the theorem. The map
\be
z_1 \mapsto \sabs{z_1}^2 +
E(z_1,0,\bar{z}_1,0,0)\,,
\ee
has a strict minimum at the origin, and so does 
\be
z_1 \mapsto \sabs{z_1}^2 + \sum_{k=2}^n \epsilon_k \sabs{z_k}^2 +
E(z_1,z',\bar{z}_1,\bar{z}',\Re w) - \Im w
\quad \text{
for small $z'$, $w$.}
\ee
We find an analytic disc $\Delta$
``attached'' to $K \subset M$
(i.e., $\partial \Delta \subset K$):
\be
\begin{adjustbox}{valign=c}
\scalebox{1}{
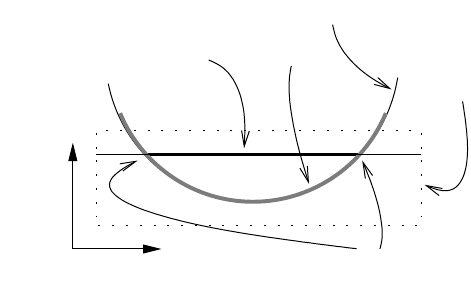
}
\end{adjustbox}
\ee
One can fill a one-sided
neighborhood by such discs.

The next step is to apply Baouendi--Tr{\`e}ves to find $p_{\ell}$
that approximate $f$ uniformly on $K$. The sequence
$\{ p_\ell \}$ is (uniformly) Cauchy on $\partial \Delta$ for each disc.
By the maximum principle, $\{ p_\ell \}$ is (uniformly) Cauchy
on $\Delta$. This implies that $\{ p_\ell \}$ is (uniformly) Cauchy on $U_- \cup K$, and it follows that $\{ p_\ell \}$ converges to a holomorphic function on $U_-$
continuous up to the boundary.
To see (iii), extend to one side, then use the Tomato can principle or Kontinuit\"atssatz to
extend to the other side.

\begin{mdexample}
Every CR function on $\Im w = \sabs{z_1}^2-\sabs{z_2}^2$ extends to an
entire holomorphic function on $\C^3$ and hence must be real-analytic.
\end{mdexample}

\begin{mdexample}
Every CR function on $\Im w = \sabs{z_1}^2+\sabs{z_2}^2$ extends to
the set $\Im w \geq \sabs{z_1}^2+\sabs{z_2}^2$, but not necessarily below.
\end{mdexample}

\begin{mdexample}
There exist CR functions on $\Im w = 0$ that extend to neither side.
\end{mdexample}

\begin{mdremark}
These ideas led Lewy to find an example of an unsolvable
PDE.
\end{mdremark}

Another application is a special case of the following theorem.
\begin{mtheorem}[Hartogs--Bochner]
Suppose $U \subset \C^n$, $n \geq 2$, is a bounded open set with a smooth boundary, and $f \colon \partial U \to \C$ is
a CR function.
Then there exists a continuous $F \colon \widebar{U} \to \C$
holomorphic in $U$
such that $F|_{\partial U} = f$.
\end{mtheorem}
The special case is if we have at least one positive Levi eigenvalue
at each point and if we can extend through compacts. This is something we will study in the next lecture.

\begin{mdremark}
Neither Hartogs nor Bochner proved this; it was proved by
Martinelli.
\end{mdremark}

\begin{mdexample}
Every CR function on $S^{2n-1} \subset \C^n$, $n \geq 2$,
is the boundary value
of a continuous $F \colon \overline{\bB_n} \to \C$ that is holomorphic in
$\bB_n$.
\end{mdexample}

\begin{mdexample}
The function $\bar{z}$ on $S^1 \subset \C$ is not the boundary value
of a holomorphic function on the disc; it would have a pole.
\end{mdexample}

\begin{mdexample}
Similarly, this is not true in general if $U$ is unbounded. If
$U = \D \times \C \subset \C^2$,
then $\bar{z}_1$ is a CR function but does not extend inside for the same
reason.
\end{mdexample}

\section[The \texorpdfstring{$\bar\partial$}{del-bar}-problem]{The \texorpdfstring{$\bar\partial$}{del-bar}-problem\label{sec:d-bar-problem}\\
\normalfont{\textit{Sean Curry}}}

\subsection{Levi form}\label{subsec:Levi-revisited}

Let us return to the notions of pseudoconvexity that we started to explore in Sec.~\ref{sec:Sean-lecture-2}.

Recall that in the previous lectures we considered domains of $\C^n$ that can be written as $U = \{ r<0 \}$, where $r = 0$ on the boundary $M = \partial U$. We considered the complexification of the tangent bundle, which takes the form
\be
\C TM = T^{(1,0)}M \oplus T^{(0,1)}M \oplus \C T\, ,
\ee
where $T^{(1,0)}_pM \oplus T^{(0,1)}_pM$ is the natural decomposition of $\C H_p$ into $i$ and $-i$ egenspaces of $J$ at each $p\in M$ and $T$ is a choice of real tangent vector field that spans a complementary subspace\footnote{This is sometimes referred to as ``the bad direction.'' Note, however, that the article ``the'' is not totally justified in that, except for in certain cases of domains with special symmetry, there is no biholomorphically invariant way to choose this complementary subspace.} to $H$ at each point. 
This decomposition of the complexified tangent space leads to a range of possible conditions on the Hessian of a defining function.
As previously, let us write the Hessian along the tangent bundle as
\be
\text{Hess}(r) \big|_{TM} = \begin{pmatrix}
r_{\alpha\beta} & r_{\alpha \bar{\beta}} & r_{\alpha 0}\\
r_{\bar{\alpha}\beta} & r_{\bar{\alpha}\bar{\beta}} & r_{\bar{\alpha}0}\\
r_{0\beta} & r_{0\bar{\beta}} & r_{00}
\end{pmatrix} .
\ee
Then different the levels of convexity mentioned previously can be stated, informally, as follows:
\begin{align}
\label{ineq:convexity}\text{Convexity:}& \qquad \text{Hess}(r)|_{TM} \geq 0\, ,\\
\label{ineq:C-convexity} \C\text{-linear convexity:}& \qquad \begin{pmatrix}
r_{\alpha\beta} & r_{\alpha \bar{\beta}} \\
r_{\bar{\alpha}\beta} & r_{\bar{\alpha}\bar{\beta}} \\
\end{pmatrix} \geq 0\, ,\\
\label{ineq:psi-convexity}\text{Pseudoconvexity:}& \qquad r_{\alpha \bar{\beta}} \geq 0\, .
\end{align}
More precisely, by \eqref{ineq:convexity} we mean that the Hessian of $r$ is positive semidefinite as a quadratic form at each point $p\in M$ when restricted to real tangent vectors to $M$, by \eqref{ineq:C-convexity} we mean that the Hessian of $r$ is positive semidefinite as a quadratic form at each point $p\in M$ when restricted to $H_p\subseteq T_p M$ (strictly speaking, since $T^{(1,0)}M \oplus T^{(0,1)}M = \C H$, the notation in \eqref{ineq:C-convexity} translates to ``$\text{Hess}(r)|_{\C H}\geq 0$,''  but what we really mean is that $\text{Hess}(r)|_{H}\geq 0$, i.e.,  the corresponding \textit{real} quadratic form is positive semidefinite), and by \eqref{ineq:psi-convexity}  we mean that the Hessian of $r$ defines a positive semidefinite Hermitian form on $T_p^{(1,0)}M$ for each $p\in M$, that is, at each point $p$ we have $\text{Hess}(r)(X_p,\overline{X_p}) \geq 0$ for all $X_p\in T^{(1,0)}_p M$.

Let us focus on pseudoconvexity, which is the weakest requirement (and the only one that is biholomorphically invariant). In practice, it amounts to checking that
\be
\sum_{j,k=1}^{n} a_j \bar{a}_k \frac{\partial^2 r}{\partial z_j \partial \bar{z}_k} \geq 0 \qquad \text{for all} \quad X_p = \sum_{j=1}^{n} a_j \frac{\partial r}{\partial z_j} \bigg|_{p} \in T^{(1,0)}_p M\, . 
\ee
The Hermitian quadratic form on the left-hand side is called the \emph{Levi form} of $r$ (or of $M = \partial U$):
\be
\mathrm{Levi}(r)|_p (X_p,Y_p) = \text{Hess}(r)|_p (X_p,\overline{Y_p}) \qquad \text{for} \quad X_p, Y_p \in T^{(1,0)}_pM.
\ee
Pseudoconvexity just says that that $\mathrm{Levi}(r) \geq 0$ as a Hermitian form. The signature of the Levi form is invariant and well-defined (it is independent of the choice of defining function and is invariant under local biholomorphic changes of coordinates). This is a consequence of the fact that the Levi form is actually invariant as a line bundle-valued Hermitian form on $T^{(1,0)}M$.

\begin{mdexample}\label{ex:Heisenberg-in-C3}
Consider $\C^3$ with complex Cartesian coordinates $(z_1,z_2,w)$. The domain defined by $\Im \, w > |z_1|^2 + |z_2|^2$ has boundary $M$ defined by $\Im \, w = |z_1|^2 + |z_2|^2$. A natural choice of defining function for our domain is $r= |z_1|^2 + |z_2|^2 - \Im \, w$.
A frame for the holomorphic tangent bundle $T^{(1,0)}M$ is given by
\be
Z_1 = \frac{\partial}{\partial z_1} + 2i\bar{z}_1 \frac{\partial}{\partial w},\qquad Z_2 = \frac{\partial}{\partial z_2} + 2i\bar{z}_2 \frac{\partial}{\partial w}.
\ee
To see this it suffices to note that $Z_{1}$ and $Z_2$ are pointwise linearly independent, lie in the span of the holomorphic vector fields $\frac{\partial}{\partial z_1}$, $\frac{\partial}{\partial z_2}$ and $\frac{\partial}{\partial w}$ on $\C^3$ and satisfy $Z_{\alpha}r=0$ for $\alpha =1,2$ (these vector fields happen to be defined and satisfy these conditions on $\C^3$ but we really only need these things to be satisfied on $M$). A frame for $T^{(1,0)}M$ is given by $Z_{\bar{1}}= \overline{Z_1}$, $Z_{\bar{2}}= \overline{Z_2}$. We wish to compute the Levi form components $\text{Hess}(r) (Z_{\alpha},Z_{\bar{\beta}})$. Note that for for general (real or complex) tangent vector fields $X$ and $Y$ on $\C^3$, computing $\text{Hess}(r) (X,\overline{Y})$ is not the same as computing $X(Yr)$ or $Y(Xr)$ since $\text{Hess}(r) (X,\overline{Y})$ is tensorial in $X$ and $Y$ whereas the expression $X(Yr)$ involves the $X$ derivatives of the coefficients of $Y$ and $Y(Xr)$ involves the $X$ derivatives of the coefficients of $Y$; the correct formula, for arbitrary vector fields $X$ and $Y$, is
\be
\text{Hess}(r) (X,Y) = Y(Xr) - (\nabla_Y X)r,
\ee
where $\nabla_Y X$ is the vector field obtained from $X$ by differentiating its components in the direction of $Y$ and all vector fields on the right hand side of the above display are acting on the functions to their right as directional derivative operators. Using the above displayed formula it is straightforward to compute that the Levi form of $r$ has components
\be
\text{Hess}(r) (Z_{\alpha}, Z_{\bar{\beta}}) = 2\delta_{\alpha\bar{\beta}}\,,
\ee
at at any point $p\in M$ (in more general examples the components of the Levi form expressed in a general frame will be functions rather than constants; but, if the Levi form is positive definite at some point, then by Gram-Schmidt it is possible to find a frame in a neighborhood of that point such that the Levi form has components $\delta_{\alpha\bar{\beta}}$, or $2\delta_{\alpha\bar{\beta}}$ if preferred, at all points in that neighborhood).

\end{mdexample}

\begin{mdexample}\label{ex:indefinite-Levi}
Again, consider $\C^3$ with coordinates $(z_1,z_2,w)$. The domain defined by $\Im \, w > |z_1|^2 - |z_2|^2$ has boundary $M$ defined by $\Im \, w = |z_1|^2 - |z_2|^2$. 
A frame for the holomorphic tangent bundle $T^{(1,0)}M$ is given by
\be
Z_1 = \frac{\partial}{\partial z_1} + 2i\bar{z}_1 \frac{\partial}{\partial w},\qquad Z_2 = \frac{\partial}{\partial z_2} - 2i\bar{z}_2 \frac{\partial}{\partial w},
\ee
and a frame for $T^{(1,0)}M$ is given by $Z_{\bar{1}}= \overline{Z_1}$, $Z_{\bar{2}}= \overline{Z_2}$. Computing as in the previous example, the Levi form of the defining function $r= |z_1|^2 - |z_2|^2 - \Im \, w$ is represented in the frame $Z_1, Z_2$ by the matrix $\text{diag}\{2,-2\}$.

\end{mdexample}
\begin{mdexample}\label{ex:levi-flat}
Consider $\C^n$ with the coordinates $(z,w)=(z_1,\ldots, z_{n-1},w)$. The domain defined by $\Im \, w >0$ has boundary $M$ defined by $\Im \, w = 0$. 
A frame for the holomorphic tangent bundle $T^{(1,0)}M$ is given by the vector fields $Z_{\alpha} = \frac{\partial}{\partial z_{\alpha}}$ for $\alpha = 1,\ldots , n-1$. The defining function $r=-\Im \, w$ is linear, meaning that $\text{Hess}(r) \equiv 0$, and hence the Levi form of this domain is identically zero; we say that its boundary is \textit{Levi flat}. Note that the boundary $M=\mathbb{C}^{n-1}\times \R$, being Levi flat, is pseudoconvex from both sides (meaning that both $\Im \, w >0$ and $\Im \, w <0$ are pseudoconvex domains).

\end{mdexample}

The hypersurfaces $\Im\, w = \epsilon_1 |z_1|^2 + \cdots + \epsilon_{n-1}|z_{n-1}|^2$, where $\epsilon_k \in \{1,-1, 0\}$, are local models for general hypersurfaces (when viewed up to local biholomorphism) in the following sense (Lem.~ 2.3.9 in \cite{lebl2019tasty}):
\begin{mdproposition} \label{prop:basic-normal-form}
Let $M \subset \C^n$ be a smooth real hypersurface and $p \in M$.  Then there
exists a local biholomorphic change of coordinates taking $p$ to $0$
and $M$ to the hypersurface given by
\be
\Im w = \sum_{k=1}^\alpha \sabs{z_k}^2 - \sum_{k=\alpha+1}^{\alpha+\beta}
\sabs{z_k}^2 +
E(z,\bar{z},\Re w) ,
\ee
where $E$ is $O(3)$ at the origin.
Here $\alpha$ is the number of positive eigenvalues of the Levi form at $p$,
$\beta$ is the number of negative eigenvalues and $\alpha+\beta \leq n-1$.
\end{mdproposition}

By constructing a suitable Hartogs figure (or ``tomato can'') it is easy to show that holomorphic functions on $\Im \, w > |z_1|^2-|z_2|^2$ can be extended to a neighborhood of the origin (in fact, to all of $\C^3$, but we focus here on local extendability). Combining this observation (and its natural generalization to higher dimensions) with Proposition \ref{prop:basic-normal-form} one readily obtains the following theorem (Thm.~ 2.3.11 in \cite{lebl2019tasty}):
\begin{mtheorem}[Tomato can principle\index{tomato can principle}] \label{thm:tomatocan}
Suppose
$U \subset \C^n$ is an open set with smooth boundary and
the Levi form has a negative eigenvalue at $p \in
\partial U$.
Then every holomorphic function on $U$
extends to a neighborhood of $p$.
In particular, $U$ is not
a domain of holomorphy.
\end{mtheorem}
The following figure illustrates the idea behind the proof, where here we have applied Proposition \ref{prop:basic-normal-form} and permuted the $z_k$ variables so that $z_1$ is a direction in which the Levi form is negative at $p$. 

\be
\adjustbox{valign=c,scale={0.9}{0.9}}{\tikzset{every picture/.style={line width=0.75pt}} 
\begin{tikzpicture}[x=0.75pt,y=0.75pt,yscale=-1,xscale=1]
\draw  [draw opacity=0][fill=RoyalBlue  ,fill opacity=0.2 ] (271,101) -- (343.63,101) -- (343.63,161) -- (271,161) -- cycle ;
\draw  [draw opacity=0][fill=RoyalBlue  ,fill opacity=0.2 ] (198.38,101) -- (271,101) -- (271,161) -- (198.38,161) -- cycle ;
\draw[->]     (198.38,101) -- (343.63,101) node[right]{$z_1$} ;
\draw  [draw opacity=0][fill=Maroon  ,fill opacity=0.2 ] (228,55) -- (251,55) -- (251,106) -- (228,106) -- cycle ;
\draw  [draw opacity=0][fill=Maroon  ,fill opacity=0.2 ] (251,92.42) -- (251,69.42) -- (291.01,69.42) -- (291.01,92.42) -- cycle ;
\draw  [draw opacity=0][fill=white  ,fill opacity=1 ] (232,160.71) .. controls (232,160.71) and (232,160.71) .. (232,160.71) .. controls (232,127.97) and (249.69,101.42) .. (271.5,101.42) .. controls (293.32,101.42) and (311.01,127.97) .. (311.01,160.71) -- (271.5,160.71) -- cycle ; \draw   (232,160.71) .. controls (232,160.71) and (232,160.71) .. (232,160.71) .. controls (232,127.97) and (249.69,101.42) .. (271.5,101.42) .. controls (293.32,101.42) and (311.01,127.97) .. (311.01,160.71) ;  
\draw  [draw opacity=0][fill=Maroon  ,fill opacity=0.2 ] (291,55) -- (314,55) -- (314,106) -- (291,106) -- cycle ;
\draw  [draw opacity=0][fill=black  ,fill opacity=1 ] (268.38,101) .. controls (268.38,99.55) and (269.55,98.38) .. (271,98.38) .. controls (272.45,98.38) and (273.63,99.55) .. (273.63,101) .. controls (273.63,102.45) and (272.45,103.63) .. (271,103.63) .. controls (269.55,103.63) and (268.38,102.45) .. (268.38,101) -- cycle ;
\draw[<-]    (271,41) -- (271,161) node[pos=0,right]{$w$};
\draw[<-,color=Maroon]    (307,83) -- (364,70) node[pos=1,right]{Hartog's figure};
\draw[<-]    (276,109) -- (285,132) ;
\draw (204,57.4) node [anchor=north west][inner sep=0.75pt]    {$U$};
\draw (280,137.4) node [anchor=north west][inner sep=0.75pt]    {$p$};
\end{tikzpicture}}
\ee
From the above theorem it follows that a domain of holomorphy must be pseudoconvex (since this is equivalent to the eigenvalues of the Levi form being nonnegative at every point).

\subsection{Equivalences}\label{subsec:equivalences}

We conclude our discussion of domains of holomorphy in $\C^n$ with the following theorem, giving a characterization of such domains in terms of pseudoconvexity (and some final examples). At this point we allow for domains with non-smooth boundary, with the caveat that we must then work with a different notion of pseudoconvexity (due to Hartogs) that makes sense when the boundary is not smooth and is equivalent to the (Levi) definition we have been using in the case of smooth boundary. A longer list of conditions equivalent to being a domain of holomorphy can be found in \cite{lebl2019tasty} (see also \cite{hormander1973introduction}).

\begin{mtheorem}\label{thm:equivalences}
Let $U \subseteq \C^n$ be a domain. Then the following are equivalent:
\begin{enumerate}[label=(\roman*)]
\item $U$ is a domain of holomorphy;
\item $U$ is Levi pseudoconvex (provided $\partial U$ is smooth);
\item $U$ is Hartogs pseudoconvex, i.e., there is a continuous exhaustion function that is plurisubharmonic.
\end{enumerate}
\end{mtheorem}
Here, an exhaustion function is one that goes to infinity at the boundary. Plurisubharmonic means subharmonic on each complex line (i.e.,  on each one dimensional complex affine subspace of $\C^n$). It is essentially a positivity condition on the complex Hessian; morally it means ``$\partial^2 f / \partial z_j \partial \bar{z}_k \geq 0$'' (it means precisely this if $f$ is twice continuously differentiable, but the condition still makes sense if $f$ is merely continuous). 

We have only proved that (i) implies (ii) in the case where $\partial U$ is smooth. That the notions of Levi pseudoconvexity and Hartogs pseudoconvexity are equivalent when $\partial U$ is smooth is proved, e.g., in \cite[Thm.~ 2.5.8]{lebl2019tasty}. (This equivalence allows us to say that a domain is \textit{pseudoconvex} without needing to specify which of the two definitions we are using.) That (i) implies (iii) is typically seen by showing that the domains of holomorphy are ``holomorphically convex'' (see \cite[Thm. 2.6.3]{lebl2019tasty} or \cite[Thm. 2.5.4 \& 2.5.5]{hormander1973introduction}) and hence that minus the logarithm of the distance to the boundary is purisubharmonic (see \cite[pp. 44-45]{hormander1973introduction}). Proving that a pseudoconvex domain is a domain of holomorphy is known as the Levi problem (as it was posed, at least for the smooth boundary case, by Levi in 1911) and was solved in the early 1950s due, principally, to the work of K.\ Oka. This problem played an important role in the development of analysis in several complex variables. Today, the solution to the Levi problem is most commonly presented using the techniques for the $\overline{\partial}$-problem developed by Kohn, H\"ormander and Andreotti-Vesentini (and others) in the mid 1960s; see \cite{hormander1973introduction}.

Although the prototypical example of a pseudoconvex domain in $\C^n$ would be either the unit ball or the unit polydisc (both are convex, and hence pseudoconvex), in physics-related problems it seems that one might more commonly encounter domains with Levi flat boundaries, such as in the following example.

\begin{mdexample}
An example that one might see in the physics literature is the domain $U = \C^n \setminus (\mathbb{R}_{\leq 0} \times \C^{n-1})  \subseteq \C^n$ (imagine having a holomorphic function $f(z_1,z_2\ldots, z_n)$ that is entire in each of the variables $z_2, \ldots, z_n$  but has a natural boundary along the negative real axis in the $z_1$ variable). Note that the boundary is then $M= \mathbb{R}_{\leq 0} \times \C^{n-1}$, which (up to a permutation of the coordinates) is a part of the boundary considered in Ex.~\ref{ex:levi-flat} and so its interior $\mathbb{R}_{\leq 0} \times \C^{n-1}$ is Levi flat (pseudoconvex from both sides). That the domain $U$ is pseudoconvex can be seen as coming from the fact that all domains in $\C$ are pseudoconvex, since $U$ is the product of the cut plane $\C\setminus \mathbb{R}_{\leq 0}$ with $\C^{n-1}$ and the product of pseudoconvex domains is pseudoconvex (i.e.,  a domain of holomorphy). 
\be
\hspace{-0.7cm}
\adjustbox{valign=c,scale={0.9}{0.9}}{\tikzset{every picture/.style={line width=0.75pt}} 
\begin{tikzpicture}[x=0.75pt,y=0.75pt,yscale=-1,xscale=1]
\draw   (116.3,130) -- (215,130) -- (172.7,170) -- (74,170) -- cycle ;
\draw   (295.3,129) -- (394,129) -- (351.7,169) -- (253,169) -- cycle ;
\draw[->]    (410,151) -- (471,151) ;
\draw  [fill=Gray ,fill opacity=0.2 ] (534.3,130) -- (583.65,130) -- (541.35,170) -- (492,170) -- cycle ;
\draw [line width=1.1]   (562.5,150) --  (562.5,103);
\draw [line width=1.1]   (562.5,150) -- (562.5+49.35,150);
\draw [line width=1.1]    (562.5,150) -- (562.5-21.15,150+20) ;
\draw [line width=1.5]    (96,149) -- (144.5,150) ;
\draw[<-]    (117,155) -- (101,185) ;
\draw (222,143.4) node [anchor=north west][inner sep=0.75pt]    {$\times $};
\draw (326,103.4) node [anchor=north west][inner sep=0.75pt]    {$\C^{n-1}$};
\draw (408,125.4) node [anchor=north west][inner sep=0.75pt]    {Levi flat};
\draw (69,185.4) node [anchor=north west][inner sep=0.75pt]    {Natural boundary};
\draw (458,181.4) node [anchor=north west][inner sep=0.75pt]    {pseudoconvex on both sides};
\end{tikzpicture}}
\ee
\end{mdexample}

We have talked a lot about CR manifolds embedded as hypersurfaces in complex Euclidean space, but (just as we can abstract the notion of a surface in Euclidean space to get the notion of an abstract Riemannian manifold) one can also talk about abstract CR manifolds. These are $2n+1$ dimensional real smooth manifolds $M$ equipped with a complex rank $n$ subbundle $T^{(1,0)}M$ of the complexified tangent bundle $\C TM$ with the property that $T^{(1,0)}_p M\cap \overline{T^{(1,0)}_pM} = \{0\} \subseteq \C T_pM$ for all $p\in M$ and such that for any two vector fields $X,Y\in \Gamma(T^{(1,0)}M)$ the Lie bracket $[X,Y]$ also lies in $T^{(1,0)}M$. The CR structure $T^{(1,0)}M$ of any smooth real hypersurface in $\C^{n+1}$ satisfies these conditions.\footnote{Some people prefer to talk about hypersurfaces in $\C^{n+1}$ rather than in $\C^n$ so that the dimension of the hypersurface is $2n+1$ rather than $2n-1$ and the CR dimension is $n$ rather than $n-1$. Sean is doing this locally (by force of habit) for this example. There is no other significance in this change of convention.} Here is an example of an abstractly defined CR manifold (which will turn out to be equivalent to a familiar hypersurface in $\C^{n+1}$):

\begin{mdexample}
Consider the abstract Heisenberg CR structure on $\mathbb{H}_n = \C^n \times \R$ (not viewed as a subset of $\C^{n+1}$) with coordinates $(z_1, \ldots, z_n, t)$. We have
\be\label{ex:frame-Heis}
T^{1,0} \mathbb{H}_n = \text{span} \{Z_1, Z_2, \ldots, Z_n \}\, ,
\ee
where
\be\label{eq:abstract-Heis-Zj}
Z_j = \frac{\partial}{\partial z_j} + i \bar{z}_j \frac{\partial}{\partial t}\, .
\ee
Let us write $Z_{\bar{j}} = \overline{Z_j}$ and $T = \frac{\partial}{\partial t}$. Then, one can show that
\begin{align}
[T , Z_j] = 0 \, ,\qquad[Z_j , Z_k] = 0 \, ,\qquad [Z_j, \bar{Z}_k ] = -2i \delta_{j\bar{k}} T\, .
\end{align}
Here, $2\delta_{j\bar{k}}$ are the abstract Levi form components (the Levi form can be defined abstractly as $i[X,\overline{Y}] \mod T^{(1,0)}\oplus T^{(1,0)}$ for sections $X$ and $Y$ of $T^{(1,0)}$; this expression is easily checked to be tensorial in $X$ and $Y$ and gives a well-defined line bundle valued Hermitian form on $T^{(1,0)}$, where the line bundle is the complex tangent bundle mod $T^{(1,0)}\oplus T^{(1,0)}$; one needs to trivialize this line bundle to get an ordinary Hermitian form on  $T^{(1,0)}$ and this is done using the choice of $T$ which is analogous to the choice of defining function $r$ in the embedded case). The Heisenberg CR structure gets its name from these commutation relations and the (corresponding) fact that there is a natural Heisenberg group structure on $\mathbb{H}_n = \C^n \times \R$ with respect to which the frame $Z_1, Z_2, \ldots, Z_n$ is left invariant.

The map
\be \label{ex:CR-embedding-Heis}
(z_1, \ldots, z_n, t) \mapsto (z_1, \ldots, z_n,\, \underbracket[0.4pt]{t + i \lVert z \rVert^2}_{w})\,,
\ee
embeds the abstract CR manifold $(\mathbb{H}_n, T^{(1,0)}\mathbb{H}_n)$ as the boundary $M$ of $\Omega_0=\{ \Im w > \lVert z \rVert^2 \}$; this map is a \textit{CR embedding} in the sense that it identifies the abstract CR structure $(\mathbb{H}_n, T^{(1,0)}\mathbb{H}_n)$ with the embedded one $(M,T^{(1,0)}M)$, meaning that the map is a smooth embedding with image $M$ and the differential of the embedding at each point $p\in \mathbb{H}_n$ induces an isomorphism between $T^{(1,0)}_p\mathbb{H}_n$ and  $T^{(1,0)}_p M$.  A smooth embedding is a CR embedding if and only if its component functions satisfy the \textit{tangential Cauchy-Riemann equations}, meaning that the functions are in the kernel of some, equivalently any, local frame $Z_{\bar{1}},\ldots, Z_{\bar{n}}$ for the antiholomorphic tangent bundle of the abstract CR manifold in a neighborhood of each point. In the case of our example it is easy to check that the components $z_1, \ldots, z_n,w=t + i \lVert z \rVert^2$ of the map \eqref{ex:CR-embedding-Heis} satisfy the tangential Cauchy-Riemann equations (i.e.,  lie in the kernel of the operators $Z_{\bar{1}},\ldots, Z_{\bar{n}}$) and this is the quickest way to see that the map is a CR embedding. 
\be
\adjustbox{valign=c,scale={0.9}{0.9}}{\tikzset{every picture/.style={line width=0.75pt}}    
\begin{tikzpicture}[x=0.75pt,y=0.75pt,yscale=-1,xscale=1]
\draw  [draw opacity=0] (451.01,109.8) .. controls (447,138.9) and (429.53,168.5) .. (411.96,175.95) .. controls (411.39,176.19) and (410.82,176.41) .. (410.25,176.6) -- (419.18,123.19) -- cycle ; \draw   (451.01,109.8) .. controls (447,138.9) and (429.53,168.5) .. (411.96,175.95) .. controls (411.39,176.19) and (410.82,176.41) .. (410.25,176.6) ;  
\draw  [draw opacity=0] (410.25,176.6) .. controls (407.45,177.41) and (404.59,177.54) .. (401.72,176.9) .. controls (389.33,174.13) and (380.66,157.72) .. (379.17,137) -- (408.94,124.15) -- cycle ; \draw   (410.25,176.6) .. controls (407.45,177.41) and (404.59,177.54) .. (401.72,176.9) .. controls (389.33,174.13) and (380.66,157.72) .. (379.17,137) ;  
\draw    (221.17,136) -- (379.17,137) ;
\draw    (246,176.5) -- (406,177.5) ;
\draw  [draw opacity=0] (252.25,175.6) .. controls (249.45,176.41) and (246.59,176.54) .. (243.72,175.9) .. controls (231.33,173.13) and (222.66,156.72) .. (221.17,136) -- (250.94,123.15) -- cycle ; \draw   (252.25,175.6) .. controls (249.45,176.41) and (246.59,176.54) .. (243.72,175.9) .. controls (231.33,173.13) and (222.66,156.72) .. (221.17,136) ;  
\draw    (293.01,108.8) -- (451.01,109.8) ;
\draw  [draw opacity=0] (293.01,108.8) .. controls (291.71,118.29) and (288.96,127.84) .. (285.25,136.66) -- (261.18,122.19) -- cycle ; \draw   (293.01,108.8) .. controls (291.71,118.29) and (288.96,127.84) .. (285.25,136.66) ;  
\draw  [draw opacity=0][dash pattern={on 4.5pt off 4.5pt}] (285.25,136.66) .. controls (277.59,154.93) and (265.77,170.07) .. (253.9,175.1) .. controls (253.32,175.35) and (252.75,175.57) .. (252.19,175.76) -- (261.11,122.35) -- cycle ; \draw  [dash pattern={on 4.5pt off 4.5pt}] (285.25,136.66) .. controls (277.59,154.93) and (265.77,170.07) .. (253.9,175.1) .. controls (253.32,175.35) and (252.75,175.57) .. (252.19,175.76) ;  
\draw[<-]    (330,68.67) -- (330,137.5) node[pos=0,right]{$\text{Im}(\omega)$};
\draw  [dash pattern={on 0.84pt off 2.51pt}]  (330,137.5) -- (330,160) ;
\draw  [dash pattern={on 0.84pt off 2.51pt}]  (330,160) -- (385,160.5) ;
\draw[->]    (385,160.5) -- (452,160.5) node[pos=1,right]{$\text{Re}(\omega)$};
\draw  [dash pattern={on 0.84pt off 2.51pt}]  (330,160) -- (313,175.5) ;
\draw[<-]    (270,208.5) -- (313,175.5) ;
\end{tikzpicture}}
\ee
Note that in this example, the abstract holomorphic tangent vector fields $Z_1,  \ldots, Z_n$ defined in \eqref{eq:abstract-Heis-Zj} are pushed forward by the diffeomorphism $\mathbb{H}_n\to M\subseteq \C^{n+1}$ to the vector fields
\be\label{eq:Heis-vfs}
\frac{\partial}{\partial z_1} + 2i\bar{z}_1 \frac{\partial}{\partial w}\;, 
\quad  \ldots \quad ,\,\; \frac{\partial}{\partial z_n} + 2i\bar{z}_n \frac{\partial}{\partial w}\,,
\ee
defined along $M$, where in this last display $z_1, \ldots z_n$ are the first $n$ coordinates on $\C^{n+1}$ whereas in \eqref{eq:abstract-Heis-Zj} the $z_1, \ldots z_n$ are coordinates for the $\C^n$ factor of the abstract manifold $\mathbb{H}_n = \C^n \times \R$ (so ``$\frac{\partial}{\partial z_j}$'' does not mean the same thing in \eqref{eq:abstract-Heis-Zj} as it does in \eqref{eq:Heis-vfs}; it might have been wise to use $(\zeta_1, \ldots , \zeta_n, w)$ rather than $(z_1,\ldots, z_n,w)$ to denote the coordinates on $\C^{n+1}$ but we chose not to). To see this, let $u= \Re w$ and $v=\Im w$ and note that $\frac{\partial}{\partial w} = \frac{1}{2}\left(\frac{\partial}{\partial u} - i \frac{\partial}{\partial v} \right)$. Clearly the vector field $\frac{\partial}{\partial t}$ on $\mathbb{H}_n$ is pushed forward to the vector field $\frac{\partial}{\partial u}$ on $M$. On the other hand, since $w$ also depends on $z$, the vector field $\frac{\partial}{\partial z_j}$ on $\mathbb{H}_n$ is pushed forward to the vector field $\frac{\partial}{\partial z_j} + i\bar{z}_j\frac{\partial}{\partial w} - i\bar{z}_j\frac{\partial}{\partial \bar{w}} = \frac{\partial}{\partial z_j} + \bar{z}_j \frac{\partial}{\partial v}$  (a quick way to check this is to use $(\zeta_1, \ldots , \zeta_n, w)$ instead of $(z_1,\ldots, z_n,w)$ for the coordinates on $\C^{n+1}$, so that $\zeta_j = z_j$ and $w = t+ i \lVert z \rVert^2$, and then use the chain rule to relate the various partial derivative operators, where one thinks of the map $\mathbb{H}_n\to M\subseteq \C^{n+1}$ as a parametrization of $M$ by $(z,t)$). Combining these observations establishes our claim (since $i\bar{z}_j \frac{\partial}{\partial u} + \bar{z}_j \frac{\partial}{\partial v} = 2i\bar{z}_j\frac{\partial}{\partial w}$).

Finally, note that $r = \lVert z \rVert^2 - \Im w$ is a natural choice of defining function for $\Omega_0$ and (as in Ex.~\ref{ex:Heisenberg-in-C3}) one can easily verify that the Levi form of $r$ in this frame has components $2\delta_{j\bar{k}}$. In particular, the Levi form is positive definite at every point.
\end{mdexample}

\subsection{Introduction to the \texorpdfstring{$\bar{\partial}$}{del-bar}-equation}

First of all, we need to understand what is the $\bar\partial$-equation ($\bar\partial$ is pronounced ``dee bar''). Let us introduce the notation
\be\label{eq:d-bar-def}
\bar\partial f = \left(\frac{\partial f}{\partial \bar{z}_1}, \ldots, \frac{\partial f}{\partial \bar{z}_n} \right) = f_{\bar{z}_1} \d \bar{z}_1 + \ldots + f_{\bar{z}_n} \d \bar{z}_n\, . 
\ee
It will be the most convenient to think of $\bar\partial f$ as a $(0,1)$-form, as on the right-hand side. We assume some familiarity with forms, though it will not be essential. On $U \subseteq \C^n$, the $\bar\partial$ operator on functions described above extends to a family of maps between the spaces $\wedge^{0,k}$ of $(0,k)$-forms:
\be
\wedge^{0,0} \xrightarrow{\bar\partial} \wedge^{0,1} \xrightarrow{\bar\partial} \wedge^{0,2} \xrightarrow{\bar\partial} \ldots \xrightarrow{\bar\partial} \wedge^{0,n}\, .
\ee
The space $\wedge^{0,n}$, for example, is spanned by forms proportional to $\d \bar{z}_1 \wedge \cdots \wedge \d \bar{z}_n$. We will focus our attention on the first operator $\wedge^{0,0} \xrightarrow{\bar\partial} \wedge^{0,1}$ defined as in \eqref{eq:d-bar-def}, though this necessitates some consideration of the next operator $\wedge^{0,1} \xrightarrow{\bar\partial} \wedge^{0,2}$. 

For a $(0,1)$-form $g$, we write $g = g_1 \d \bar{z}_1 + \ldots + g_n \d \bar{z}_n$. Then
\be\label{eq:del-bar-g}
\bar{\partial} g = 0 \qquad\Leftrightarrow\qquad \partial_{\bar{z}_j} g_{\bar{k}} = \partial_{\bar{z}_k} g_{\bar{j}} \quad \text{for all }j,k.
\ee
Finally, it is straightforward to show that since the second partial derivatives commute we have 
\be
\bar\partial^2 = 0\, .
\ee

We are now ready to state the $\bar\partial$-problem (in the lowest degree case): Given a $(0,1)$-form $g$ satisfying the obvious necessary condition \eqref{eq:del-bar-g}, can we solve $\bar{\partial }f = g$?

A sharper question is: can we find a solution operator? Alternatively, can we ``solve with estimates''? The last class of problems has for a long time been one of the most active areas of research in several complex variables, tracing its roots back to the work of Kohn, H\"ormander and others in the 1960s. Being able to ``solve with estimates'' is very useful. Suppose that we are trying to construct a holomorphic function $F$ (or some other ``holomorphic'' object) with certain desired properties, and we know that we can construct such a function that is smooth and close to being holomorphic (in an appropriate sense). Then we can take our approximately holomorphic function $F_{\text{approx}}$ and set $g = \bar{\partial} F_{\text{approx}}$, which measures the failure of our approximate solution to be holomorphic and should be ``small'' (and will satisfy $\bar{\partial} g = 0$ since $\bar\partial^2 = 0$).  Then, if we are able to solve the $\bar\partial$-problem with estimates, this gives a solution $f$ to $\bar{\partial }f = g$ with $f$ ``small'' (in some norm it should be bounded by some other norm of $g$, which is small). In particular, the smallness of $f$ should force $f$ to be different from, and relatively small compared to $F_{\text{approx}}$ (both solve the $\bar{\partial }$-equation with $g$ as the right-hand side, but remember that the space of solutions to this equation is very large since holomorphic functions lie in the kernel of $\bar{\partial }$); this is done by choosing appropriate function spaces/norms. Our desired holomorphic function $F$ is then $F_{\text{approx}}-f$, and the ``smallness'' of $f$ ensures that $F$ retains the desired characteristics that $F_{\text{approx}}$ had by construction.

As a simple example of this (where the only ``estimates'' we need are the local boundedness of the solution $f$) suppose that we are supposed to find a meromorphic function $F$ on $\C$ with prescribed poles at a finite number of points $p_1, \ldots, p_m$ and with $F$ having prescribed principal part of its Laurent expansion (the part involving negative powers) at each of these poles. Picking discs $\Delta_{r_1}(p_1), \ldots ,\Delta_{r_m}(p_m)$ around these points whose closures are disjoint, we may define $F_{\text{approx}}$ in each punctured disc $\Delta_{r_j}(p_j)\setminus\{p_j\}$ to be equal to the polynomial in $\frac{1}{z-p_j}$ that is prescribed as the principal part of the Laurent expansion of $F$. We may then smoothly extend $F_{\text{approx}}$ to all of $\C$ with compact support (meaning that $F_{\text{approx}}$ is zero outside of a large enough ball). 
\be
\adjustbox{valign=c,scale={0.9}{0.9}}{\tikzset{every picture/.style={line width=0.75pt}}
\begin{tikzpicture}[x=0.75pt,y=0.75pt,yscale=-1,xscale=1]
\draw  [fill={rgb, 255:red, 155; green, 155; blue, 155 }  ,fill opacity=0.2 ] (151.44,65.13) .. controls (120.42,27.07) and (367.13,46.8) .. (338.93,75) .. controls (310.74,103.19) and (221.92,107.42) .. (250.12,149.71) .. controls (278.31,192.01) and (120.42,182.14) .. (141.57,151.12) .. controls (162.71,120.11) and (182.45,103.19) .. (151.44,65.13) -- cycle ;
\draw  [dash pattern={on 4.5pt off 4.5pt}] (172,142) .. controls (172,132.61) and (179.61,125) .. (189,125) .. controls (198.39,125) and (206,132.61) .. (206,142) .. controls (206,151.39) and (198.39,159) .. (189,159) .. controls (179.61,159) and (172,151.39) .. (172,142) -- cycle ;
\draw  [dash pattern={on 4.5pt off 4.5pt}] (181,69) .. controls (181,59.61) and (188.61,52) .. (198,52) .. controls (207.39,52) and (215,59.61) .. (215,69) .. controls (215,78.39) and (207.39,86) .. (198,86) .. controls (188.61,86) and (181,78.39) .. (181,69) -- cycle ;
\draw  [dash pattern={on 4.5pt off 4.5pt}] (251,77) .. controls (251,67.61) and (258.61,60) .. (268,60) .. controls (277.39,60) and (285,67.61) .. (285,77) .. controls (285,86.39) and (277.39,94) .. (268,94) .. controls (258.61,94) and (251,86.39) .. (251,77) -- cycle ;
\draw[->]    (277,130) -- (270,88) node[pos=0,right]{make $F_{\text{approx}}$ equal to the}
node[pos=0,right,yshift=-15]{prescribed Laurent expansion}
node[pos=0,right,yshift=-30]{around each of the sings.};
\draw (180.73,134.34) node [anchor=north west][inner sep=0.75pt]    {$\times $};
\draw (189.96,60.35) node [anchor=north west][inner sep=0.75pt]    {$\times $};
\draw (260.45,69.4) node [anchor=north west][inner sep=0.75pt]    {$\times $};
\end{tikzpicture}}
\ee
Note that $g= \bar{\partial }F_{\text{approx}}$ will be zero inside each of the punctured discs, and hence can be extended by zero across the points $p_1, \ldots, p_m $ so as to be defined on all of $\C$. Thus, we have a compactly supported smooth $(0,1)$-form $g$ (that trivially satisfies $\bar{\partial }g = 0$) and if we know that there is a locally bounded solution $f$ to $\bar{\partial }f = g$, then $F=F_{\text{approx}}-f$ will be holomorphic on $\C\setminus\{p_1, \ldots, p_m\}$ and (due to the boundedness of $f$ near each point $p_j$) will have the prescribed prescribed polynomial in $\frac{1}{z-p_j}$ as the principle part of its Laurent expansion at each of the $p_j$. We will see in the discussion that follows how to find such a function $f$.

\subsection{The generalized Cauchy integral formula}

Before we get into the $\bar\partial$-problem, let us state a more general version of Cauchy’s
formula using Stokes’ theorem (really, Green’s theorem). This version is called the
Cauchy--Pompeiu integral formula. We only need the theorem for smooth functions, but,
as it is often applied in less regular contexts and is just an application of Stokes’
theorem, let us state it more generally. In applications, the boundary is often only
piecewise smooth, and again that is all we need for Stokes.
\begin{mtheorem}
In one complex variable, we have the Cauchy--Pompeiu formula:
\be
f(z) = \frac{1}{2\pi i } \int_{\partial U} \frac{f(\zeta)}{\zeta - z} \d \zeta + \frac{1}{2\pi i} \int_U \frac{\frac{\partial f}{\partial \bar{\zeta}}(\zeta)}{\zeta - z} \underbrace{\d \zeta \wedge \d \bar{\zeta}}_{-2i \d x \wedge \d y}\, .
\ee
for $z \in U$ with piecewise $C^1$ boundary $\partial U$ and $f : \bar{U} \to \C$ continuous with bounded partial derivatives in $U$.
\end{mtheorem}
The proof involves excising a small disk $\Delta_r(z)$ from $U$ and applying Stokes' theorem on $U \setminus \Delta_r(z)$.
\be
\begin{adjustbox}{valign=c}
\scalebox{1}{
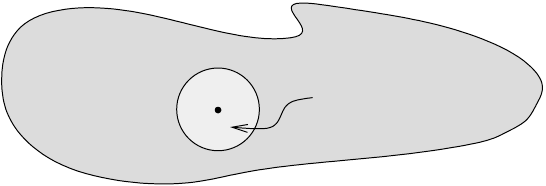
}
\end{adjustbox}
\ee

The basic idea for several complex variables is to recycle the above 1D problem but also use $\bar{\partial} g= 0$, which means that the derivatives are not independent of each other and hence allows us to tie different 1D problems together.

\subsection{Compactly supported \texorpdfstring{$\bar\partial$}{del-bar}-problem}

Let us now consider $U = \C^n$ with $n \geq 2$. Most of what we say below will also work for $n=1$.

\begin{mtheorem}
Suppose $g$ is a $(0,1)$-form on $\C^n$ with $n \geq 2$, which is smooth, compactly supported, and satisfies the integrability condition $\bar\partial g = 0$ ($\partial_{\bar{z}_j} g_k = \partial_{\bar{z}_k} g_j$ for all $j,k$) on $\C^n$. Then there exists a unique compactly supported smooth function $\psi : \C^n \to \C$ such that
\be\label{eq:del-bar-psi}
\bar\partial \psi = g\, ,
\ee
i.e., $\partial \psi / \partial \bar{z}_j = g_j$ for all $j=1,2,\ldots,n$.\\

\noindent\emph{Proof}:
We are motivated by the Cauchy--Pompeiu formula with $g_1$ replaced by $\partial \psi / \partial \bar{z}_1$ and $U$ a ``large'' ball.
We will define $\psi$ by
\be
\psi(z) = \frac{1}{2\pi i} \int_{\C} \frac{g_1(\zeta, z_2, \ldots, z_n)}{\zeta- z_1} \d\zeta \wedge \d \bar{\zeta}\, .
\ee
The aim is to differentiate the above formula and check that it satisfies \eqref{eq:del-bar-psi}. The non-trivial task is going to be to establish that the result is compactly supported.
We first change the variables $\zeta \mapsto \zeta + z_1$, which give
\be\label{eq:psi-z}
\psi(z) = \frac{1}{2\pi i} \int_{\C} \frac{g_1(\zeta + z_1, z_2, \ldots, z_n)}{\zeta} \d\zeta \wedge \d \bar{\zeta}\, .
\ee
This eliminates $z_1$ from the denominator, so we can take the derivative $\partial/\partial \bar{z}_k$ under the integral. Hence
\be\label{eq:del-bar-psi-k}
\frac{\partial \psi}{\partial \bar{z}_k}(z) = \frac{1}{2\pi i } \int_{\C} \frac{\frac{\partial g_1}{\partial \bar{z}_k}(\zeta + z_1, z_2, \ldots, z_n)}{\zeta} \d\zeta \wedge \d \bar{\zeta}\, .
\ee

On the other hand, the Cauchy--Pompeiu formula applied to $g_k$ (on $|z| \leq R$ for large $R$) gives
\be
g_k(z) = \frac{1}{2\pi i } \int_{\C} \frac{\frac{\partial g_k}{\partial \bar{z}_1}(\zeta, z_2, \ldots, z_n)}{\zeta - z_1} \d\zeta \wedge \d \bar{\zeta}\, .
\ee
We can now use the integrability condition $\partial g_1 / \partial \bar{z}_k = \partial g_k / \partial \bar{z}_1$ to write 
\be
\frac{\partial \psi}{\partial \bar{z}_k}(z) = \frac{1}{2\pi i } \int_{\C} \frac{\frac{\partial g_k}{\partial \bar{z}_1}(\zeta + z_1, z_2, \ldots, z_n)}{\zeta} \d\zeta \wedge \d \bar{\zeta}\, .
\ee
Finally, making the change of variables $\zeta \mapsto \zeta - z_1$, we find
\be
\frac{\partial \psi}{\partial \bar{z}_k}(z) = \frac{1}{2\pi i } \int_{\C} \frac{\frac{\partial g_k}{\partial \bar{z}_1}(\zeta , z_2, \ldots, z_n)}{\zeta - z_1} \d\zeta \wedge \d \bar{\zeta} = g_k(z)\, ,
\ee
by the previous formula \eqref{eq:del-bar-psi-k}.

So far, the proof works for $n \geq 1$. It remains to establish that $\psi$ has a compact support, which requires $n \geq 2$. Consider the picture:
\be
\begin{adjustbox}{valign=c}
\scalebox{1}{
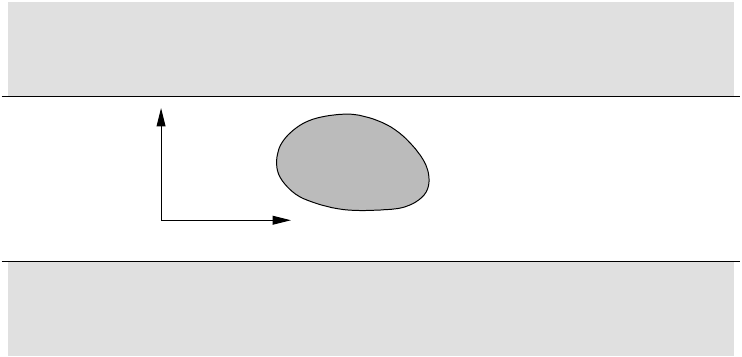
}
\end{adjustbox}
\ee
From the formula for $\psi$, $\psi \equiv 0$ in the regions where at least one of $z_2, \ldots, z_n$ is large. However, by the identity theorem, this forces $\psi \equiv 0$ in the unbounded component of $\C^n \setminus \mathrm{supp}(g)$.
Thus, $\psi$ has compact support. \hfill$\square$
\end{mtheorem}

\subsection{The general Hartogs phenomenon}

We can now prove the general Hartogs phenomenon as an application of the solution
of the compactly supported inhomogeneous $\bar\partial$-problem. We proved special versions
of this phenomenon using Hartogs figures before. 

\begin{mtheorem}[Hartogs phenomenon]
Let $U \subseteq \C^n$ be a domain with $n \geq 2$, and let $K \subseteq U$ be a relatively compact subset of $U$ such that $U \setminus K$ is connected.
\be
\begin{adjustbox}{valign=c}
\scalebox{1}{
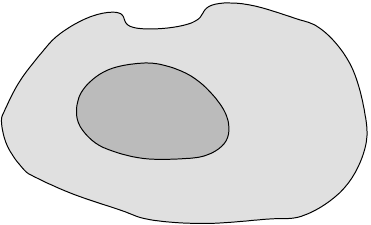
}
\end{adjustbox}
\ee
Then every holomorphic function $f \in \sO(U\setminus K)$ extends holomorphically to all of $U$.
\end{mtheorem}

The idea behind the proof is as follows. First, extend $f \in \sO(U \setminus K)$ smoothly to get the (not necessarily holomorphic) function $\tilde{f}$ on $U$, with $\bar\partial \tilde{f} $ supported in $K$ (one may need to enlarge $K$ slightly here if it is not a nice set with smooth boundary as in the picture, but this is not a problem). Then, solve $\bar\partial \psi = \bar\partial \tilde{f}$ with $\psi$ compactly supported in $K$. Finally, set $f = \tilde{f} - \psi$ on $U$.

We can finally add yet another equivalence statement to the list in Thm.~\ref{thm:equivalences}:\\
(iv) The $\bar\partial$-problem ($\bar\partial f = g$) is solvable on $(0,q)$-forms for $1 \leq q \leq n-1$.

This discussion illustrates the general idea of how one uses the solvability of the $\bar\partial$-problem.
There are a whole myriad of other setups where we can solve the $\bar\partial$-problem. For example, it is solvable on polydiscs, balls, and more generally on domains of holomorphy, and so in these cases one is naturally lead to ask about ``solvability with estimates'' (the classic references are \cite{FollandKohn,hormander1973introduction}, more recent references include \cite{ChenShaw,Straube,Ohsawa}). Solvability of the $\bar\partial$-problem also plays a key role in complex geometry; see, e.g., \cite{Demailly-book}. However, for the purposes of these lectures, we will stop here.

\newpage
\bibliographystyle{jhep}
\bibliography{references}

\providecommand{\href}[2]{#2}\begingroup\raggedright\begin{thebibliography}{10}

\bibitem{RecordsBook}
N.~Arkani-Hamed, P.~Benincasa, S.~Caron-Huot, M.~Correia, S.~Curry, M.~Giroux,
  F.~M. Haehl, H.~S. Hannesdottir, M.~T. Hansen, A.~Hebbar, G.~Isabella,
  J.~Lebl, M.~H.~G. Lee, S.~Mizera, E.~Pajer, C.~Pasiecznik, E.~Passemar,
  M.~Rangamani, B.~C. van Rees, F.~Vazão, A.~M. Wolz and Z.~Zhou,
  \emph{{Records from the S-Matrix Marathon: Selected Topics on Scattering
  Amplitudes}}. Springer Lecture Notes in Physics, 2025.

\bibitem{lebl2019tasty}
J.~Lebl, \emph{Tasty bits of several complex variables}. Independently
  published: \url{https://www.jirka.org/scv/}, 2023.

\bibitem{Fefferman1974}
C.~Fefferman, \emph{The {B}ergman kernel and biholomorphic mappings of
  pseudoconvex domains},
  \href{https://doi.org/10.1007/BF01406845}{\emph{Invent. Math.} {\bfseries 26}
  (1974) 1}.

\bibitem{Fefferman1976}
C.~L. Fefferman, \emph{Monge-{A}mp\`ere equations, the {B}ergman kernel, and
  geometry of pseudoconvex domains},
  \href{https://doi.org/10.2307/1970945}{\emph{Ann. of Math. (2)} {\bfseries
  103} (1976) 395}.

\bibitem{Fefferman1979}
C.~Fefferman, \emph{Parabolic invariant theory in complex analysis},
  \href{https://doi.org/10.1016/0001-8708(79)90025-2}{\emph{Adv. in Math.}
  {\bfseries 31} (1979) 131}.

\bibitem{FeffermanGraham1985}
C.~Fefferman and C.~R. Graham, \emph{Conformal invariants},  in \emph{\'Elie
  Cartan et les math\'ematiques d'aujourd'hui - Lyon, 25-29 juin 1984},
  pp.~95--116.
\newblock Soci\'et\'e math\'ematique de France, 1985.

\bibitem{FeffermanGraham2012}
C.~Fefferman and C.~R. Graham, \emph{The ambient metric}, vol.~178 of
  \emph{Annals of Mathematics Studies}. Princeton University Press, Princeton,
  NJ, 2012.

\bibitem{RobinsonTrautman1961}
I.~Robinson and A.~Trautman, \emph{Some spherical gravitational waves in
  general relativity},
  \href{https://doi.org/10.1098/rspa.1962.0036}{\emph{Proc. Roy. Soc. London
  Ser. A} {\bfseries 265} (1961/62) 463}.

\bibitem{Kerr1963}
R.~P. Kerr, \emph{Gravitational field of a spinning mass as an example of
  algebraically special metrics},
  \href{https://doi.org/10.1103/PhysRevLett.11.237}{\emph{Phys. Rev. Lett.}
  {\bfseries 11} (1963) 237}.

\bibitem{PenroseRindler-II}
R.~Penrose and W.~Rindler, \emph{Spinors and space-time. {V}ol. 2}, Cambridge
  Monographs on Mathematical Physics. Cambridge University Press, Cambridge,
  second~ed., 1988.

\bibitem{HillLewandowskiNurowski2008}
C.~D. Hill, J.~Lewandowski and P.~Nurowski, \emph{Einstein's equations and the
  embedding of 3-dimensional {CR} manifolds},
  \href{https://doi.org/10.1512/iumj.2008.57.3473}{\emph{Indiana Univ. Math.
  J.} {\bfseries 57} (2008) 3131}.

\bibitem{hormander1973introduction}
L.~Hormander, \emph{An introduction to complex analysis in several variables}.
  Elsevier, 1973.

\bibitem{KohnNirenberg1973}
J.~J. Kohn and L.~Nirenberg, \emph{A pseudo-convex domain not admitting a
  holomorphic support function},
  \href{https://doi.org/10.1007/BF01428194}{\emph{Math. Ann.} {\bfseries 201}
  (1973) 265}.

\bibitem{Kolar2010}
M.~Kol\'{a}\v{r}, \emph{Higher order invariants of {L}evi degenerate
  hypersurfaces}, \href{https://doi.org/10.4310/PAMQ.2010.v6.n4.a5}{\emph{Pure
  Appl. Math. Q.} {\bfseries 6} (2010) 1035}.

\bibitem{FollandKohn}
G.~B. Folland and J.~J. Kohn, \emph{The {N}eumann problem for the
  {C}auchy-{R}iemann complex}, vol.~No. 75 of \emph{Annals of Mathematics
  Studies}. Princeton University Press, Princeton, NJ; University of Tokyo
  Press, Tokyo, 1972.

\bibitem{ChenShaw}
S.-C. Chen and M.-C. Shaw, \emph{Partial differential equations in several
  complex variables}, vol.~19 of \emph{AMS/IP Studies in Advanced Mathematics}.
  American Mathematical Society, Providence, RI; International Press, Boston,
  MA, 2001, \href{https://doi.org/10.1090/amsip/019}{10.1090/amsip/019}.

\bibitem{Straube}
E.~J. Straube, \emph{Lectures on the {$\mathscr{L}^2$}-{S}obolev theory of the
  {$\overline{\partial}$}-{N}eumann problem}, ESI Lectures in Mathematics and
  Physics. European Mathematical Society (EMS), Z\"{u}rich, 2010,
  \href{https://doi.org/10.4171/076}{10.4171/076}.

\bibitem{Ohsawa}
T.~Ohsawa, \emph{{$L^2$} approaches in several complex variables}, Springer
  Monographs in Mathematics. Springer, Tokyo, second~ed., 2018,
  \href{https://doi.org/10.1007/978-4-431-56852-0}{10.1007/978-4-431-56852-0}.

\bibitem{Demailly-book}
J.-P. Demailly, \emph{Complex Analytic and Differential Geometry}. Open Access
  Online:
  \url{https://www-fourier.ujf-grenoble.fr/~demailly/manuscripts/agbook.pdf},
  2012.

\end{thebibliography}\endgroup

\end{document}